\documentclass[sigconf, authorversion]{acmart}

\usepackage{wrapfig}

\usepackage{gensymb}
\usepackage{subcaption}
\usepackage{graphicx}
\usepackage{enumitem}
\usepackage{pdfrender}
\newcommand*{\boldcheckmark}{%
  \textpdfrender{
    TextRenderingMode=FillStroke,
    LineWidth=.5pt, % half of the line width is outside the normal glyph
  }{\checkmark}%
}

\let\comment\undefined
\usepackage[final]{changes}
\AtBeginDocument{%
  \providecommand\BibTeX{{%
    \normalfont B\kern-0.5em{\scshape i\kern-0.25em b}\kern-0.8em\TeX}}}

\copyrightyear{2021}
\acmYear{2021}
\setcopyright{acmcopyright}\acmConference[UIST '21]{The 34th Annual ACM Symposium on User Interface Software and Technology}{October 10--14, 2021}{Virtual Event, USA}
\acmBooktitle{The 34th Annual ACM Symposium on User Interface Software and Technology (UIST '21), October 10--14, 2021, Virtual Event, USA}
\acmPrice{15.00}
\acmDOI{10.1145/3472749.3474768}
\acmISBN{978-1-4503-8635-7/21/10}

\acmSubmissionID{2027}

\begin{document}

% \title[NavStick]{NavStick: A Directional Surveying Tool for Blind-Accessible Video Games}

% \title[NavStick]{NavStick: Making Video Games Blind-Accessible by Giving Players the Ability to Look Around}

% \title[NavStick]{NavStick: Making Video Games Blind-Accessible by Enabling Players to Look Around}

\title{NavStick: Making Video Games Blind-Accessible via the Ability to Look Around}


% AUTHORS: 

\author{Vishnu Nair}
\affiliation{%
  \institution{Columbia University}
  \city{New York}
  \state{New York}
  \country{USA}}
\email{nair@cs.columbia.edu}

\author{Jay L. Karp}
\affiliation{%
  \institution{Columbia University}
  \city{New York}
  \state{New York}
  \country{USA}}

\author{Samuel Silverman}
\affiliation{%
  \institution{Columbia University}
  \city{New York}
  \state{New York}
  \country{USA}}

\author{Mohar Kalra}
\affiliation{%
  \institution{Columbia University}
  \city{New York}
  \state{New York}
  \country{USA}}

\author{Hollis Lehv}
\affiliation{%
  \institution{Columbia University}
  \city{New York}
  \state{New York}
  \country{USA}}

\author{Faizan Jamil}
\affiliation{%
  \institution{SUNY New Paltz}
  \city{New Paltz}
  \state{New York}
  \country{USA}}

\author{Brian A. Smith}
\affiliation{%
  \institution{Columbia University}
  \city{New York}
  \state{New York}
  \country{USA}}
\email{brian@cs.columbia.edu}

\renewcommand{\shortauthors}{Nair, et al.}

\begin{abstract}
  Video games remain largely inaccessible to visually impaired people (VIPs). Today’s blind-accessible games are highly simplified renditions of what sighted players enjoy, and they do not give VIPs the same freedom to look around and explore game worlds on their own terms. In this work, we introduce NavStick, an audio-based tool for looking around within virtual environments, with the aim of making 3D adventure video games more blind-accessible. NavStick repurposes a game controller's thumbstick to allow VIPs to survey what is around them via line-of-sight. \comment{The next two sentences were swapped to reflect the new study ordering.}In a \replaced{user}{second} study, we compare NavStick with traditional menu-based surveying for different navigation tasks \added{and find that VIPs were able to form more accurate mental maps of their environment with NavStick than with menu-based surveying}. In a\replaced{n additional exploratory}{user} study, we investigate NavStick in \replaced{the context of a representative}{different} 3D adventure game\deleted{environments, and find that it allows VIPs to complete them with a sense of fun and agency not found in current blind-accessible games}.  Our findings reveal several implications for blind-accessible games, and we close by discussing these.
\end{abstract}




%   \includegraphics[width=\textwidth]{figures/F1_IntroFigure.pdf}
%   \caption{TODO.}

\maketitle

\begin{figure}[H]
    \centering
    \includegraphics[width=0.88\linewidth]{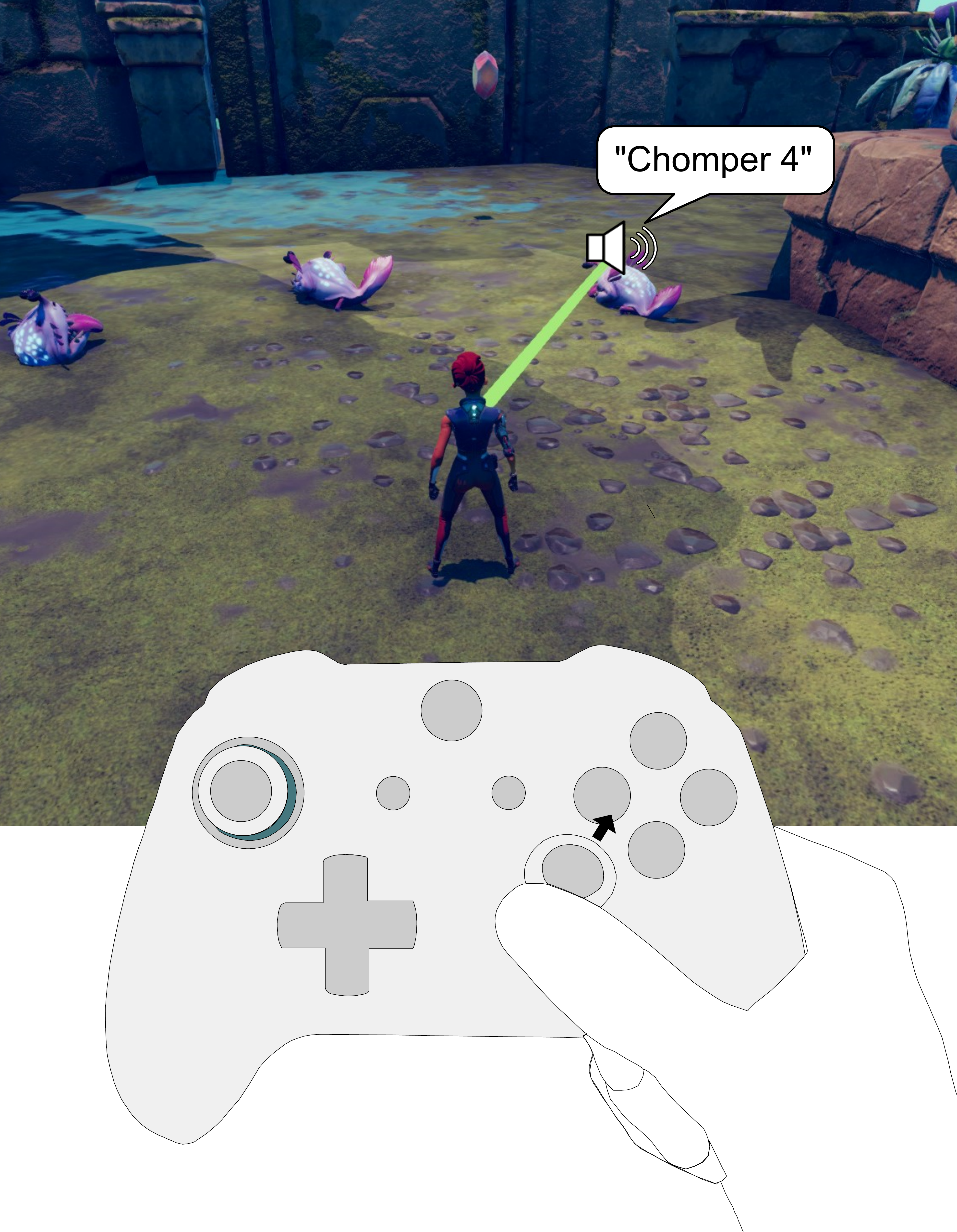}
    \caption{Overview of NavStick. \textit{Top:} A player looks around for enemy ``Chompers'' within \textit{The Explorer}, our prototype 3D adventure game. The green line represents their scrubbing direction; they are scrubbing at one o'clock and will hear the system read ``Chomper 4'' via spatialized audio. \textit{Bottom:} A player uses their game controller to look around for enemies using NavStick. Here, the player is pushing the right stick in the one o'clock direction.}
    \label{fig:navstick_overview}
\end{figure}

\section{Introduction}

Video games act as a social and emotional outlet for many people — improving mood and fostering a sense of community among players~\cite{Granic2014}. However, most mainstream 3D video games remain largely inaccessible to visually impaired people (VIPs) due to a lack of accessibility tools within them~\cite{Archambault2008, Porter2013}. As a result, VIPs are restricted to playing audio-centric blind-accessible games that are greatly simplified compared to games created for sighted players~\cite{Andrade2019, Smith2018}. These games are designed to be easier for VIPs. In order to achieve this, they confine players to 2D grid-like worlds~\cite{Matsuo2016, OutOfSightGames2017}, guide players along strict predetermined paths~\cite{NaughtyDogSonyEnt2020}, present points-of-interest (POIs) in linear menus or grids~\cite{Westin2004, PinInteractive2003a, Trewin2008}, and output announcements or sounds continuously (and often distractingly) to keep the player updated~\cite{Kaldobsky2011}.

These techniques limit the freedom that visually impaired players have in exploring the game world --- freedom that sighted players are granted by default within mainstream games. It is, thus, an important challenge to grant VIPs the same level of agency in gaming --- in part, by allowing them to play the \textit{same} mainstream 3D games that sighted players play with the \textit{same} amount of agency afforded to sighted players.

A key aspect of a sighted player's experience within a 3D video game is the ability to use the in-game camera to look around --- to find items, detect enemies, and otherwise interact with the world. Looking around gives the player agency by allowing them to understand their surrounding environment, develop a plan of action, and attempt to execute that plan~\cite{Church1999a, Church1999b, Church2006, Yuan2011}. Existing games, however, do not offer VIPs an analogue to the in-game camera that allows them to ``look around'' on their own terms. We believe that providing VIPs with such a tool is critical to providing them with a higher level of freedom and agency in exploring 3D game worlds in a \textit{self-directed way} --- that is, completely on their own terms.

Hence, we introduce \textit{NavStick}, a self-directed directional surveying tool for virtual environments that grants VIPs the ability to look around within virtual worlds. Its aim is to make existing 3D video games more blind-accessible. NavStick repurposes a game controller’s right thumbstick to allow players to survey their immediate surroundings by “scrubbing” the thumbstick in a circular fashion. As the player scrubs in different directions (their 1 o’clock, 2 o’clock, etc.), NavStick uses spatialized audio to announce what lies in those directions via line-of-sight. Figure~\ref{fig:navstick_overview} shows an example of this within a prototype 3D adventure game we created called \textit{The Explorer}.

NavStick \deleted{was co-designed alongside VIPs and }aims to replicate the ability to look around by giving users \textit{on-demand, random access} to information about their surroundings in all directions. That is, users can survey \textit{any} direction \textit{instantly} without having to traverse a list or other sequential interface. Furthermore, NavStick does not output any audio feedback unless the user is scrubbing. These properties stand in contrast to many tools in existing blind-accessible games, which have users survey their surroundings via linear menus or instructions, and which often generate unwanted and distracting sounds by outputting announcements automatically. In addition, NavStick does not require any special hardware --- it is designed to work using a standard game controller and pair of headphones. \added{We co-designed NavStick alongside a member of our research team who is a co-author on this paper and is blind. We leveraged their experience with games to identify design requirements, playing a crucial role in designing NavStick and our studies.}

\comment{The following two paras were swapped.}We evaluate NavStick through two user studies. In our \replaced{first}{second} study, we compared NavStick to menu-based surveying tools found in existing blind-accessible games by having nine VIPs perform six types of virtual navigation tasks. \deleted{(Study 1 investigates NavStick with respect to game \textit{environments}, while Study 2 investigates NavStick with respect to navigation \textit{tasks}.) }We found that NavStick allowed VIPs to form more accurate mental maps of their environment with less effort and frustration than “NavMenu,” a baseline menu-based surveying tool. Players preferred NavStick for tasks that rely on having a mental map of the surrounding environment, as well as for navigating within video games. Players also felt, however, that the ability to efficiently determine the presence of objects (as exemplified by NavMenu) and the ability to look around (as exemplified by NavStick) serve different purposes when surveying an area, and that the two tools should complement each other.

\replaced{For}{In} our \replaced{second, exploratory}{first} study, we use \textit{The Explorer} (our 3D adventure game) as a case study to investigate the ways in which NavStick can make 3D video games more blind-accessible by having seven VIPs play through the game. We found that NavStick allowed participants to experience a sense of freedom, environmental understanding, and fun not found in present-day blind-accessible games. Participants completed all portions of \textit{The Explorer}, which featured different environments that replicated the complexities of mainstream 3D adventure games. Still, some challenges in handling these complexities surfaced: NavStick had difficulty handling environmental occlusions gracefully, and players had differing opinions about NavStick's handling of moving enemies and situations involving time pressure.

% NavStick also gave VIPs a greater feeling of agency when navigating compared to NavMenu, which they found very valuable in a simple video game setting. 

We make four main contributions in this paper:
\vspace{-1mm}
\begin{enumerate}
    \item We introduce NavStick, a directional surveying tool that gives VIPs the ability to look around 3D virtual environments in a self-directed way.
    \item \comment{Contributions 2 and 3 were swapped.}We investigate the advantages and disadvantages of using NavStick compared to menu-based surveying tools (as are common in existing games) for various navigation tasks.
    \item We explore the ways in which NavStick can make \textit{existing} 3D adventure game worlds more blind-accessible.
    \item We describe NavStick's implications for future efforts on making 3D video games blind-accessible.
\end{enumerate}

\section{Related Work}

NavStick contributes to a rich history of audio navigation tools made for both video games and the real world.

\subsection{Audio Navigation Tools within Games}

The vast majority of mainstream video games are inaccessible to people with visual impairments. Hence, researchers have expressed a need for tools within games that let VIPs make decisions themselves rather than being guided and told what to do~\cite{Archambault2008, Andrade2019, Andrade2020, Smith2018}. Current means of navigating within games do not offer VIPs \textit{on-demand, random access} to information about their surroundings in all directions, which prevents VIPs from being able to play 3D games with an experience equivalent to what sighted players enjoy.

In PowerUp ~\cite{Trewin2008}, players survey the environment by scrolling through a menu of objects in the area, each of which is presented with its name, distance, and relative orientation to the player. \textit{Terraformers}~\cite{PinInteractive2003a, Westin2004}, a blind-accessible 3D adventure game, gives players an assortment of navigation tools, including a ``GPS navigator'' for determining their current position via grid coordinates and a beacon-based ``sonar'' for finding out what is directly in front of them. These tools do not, however, provide \textit{random access} to information about the player's surroundings as NavStick does. In Study~1, we recreate a level from \textit{Terraformers} to act as a testbed for comparing NavStick with menu-based surveying.

Many adventure games for VIPs, such as \textit{Shadowrine}~\cite{Matsuo2016} and \textit{A Hero’s Call}~\cite{OutOfSightGames2017}, also conform to grid- and menu-based surveying. They physically confine players to 2D grid-like worlds so that they can only move in the four cardinal directions. This simplifies the structure of these games --- POIs are situated within a single “cell” of the world, and players can query that cell to see what is present. NavStick, by contrast, does not require a grid-based world and thus enables players to look around within arbitrary game worlds as they please using line-of-sight.

Some mainstream 3D games have added accessibility features for VIPs. \textit{The Last of Us Part II}~\cite{NaughtyDog2020}, a 3D action-adventure/survival horror game released in 2020, is a prominent example. It features an ``enhanced listen mode’’ to help VIPs navigate. When triggered, audio beacons appear at the locations of nearby enemies and other POIs, but they appear all at once and may overwhelm the player if there are many targets in their vicinity. This tool also does not allow players to scan in specific directions. NavStick, by contrast, allows players to ``scan'' any direction they wish, preventing them from being overloaded with sounds and allowing them to explore the environment at their own pace. 

% \textit{The Last of Us Part II} also heavily simplifies the game for VIPs by guiding players along a ``golden path`` to progress through the storyline and allowing them to skip puzzles entirely.

\subsection{Real-World Audio Navigation Tools}

Although our focus is on virtual worlds, many sound-based methods for real-world navigation also apply within virtual worlds. Indeed, previous research has shown that game-based wayfinding tasks have ecological validity in assessing real-world navigation ability~\cite{Coutrot2019}. Existing audio navigation tools guide users along a series of waypoints connected by line segments, and they do so in two different ways. SWAN~\cite{Wilson2007}, audioGPS~\cite{Holland2002}, and Microsoft Soundscape~\cite{MicrosoftResearch2018} place acoustic ``beacons'' at each waypoint. VIPs can follow those beacons by physically turning to ``center'' the beacons' spatialized sounds in front of them, and then walking forward. NavCog~\cite{Ahmetovic2016} and NavCog3~\cite{Sato2017}, by contrast, give VIPs spoken directions to follow instead of beacons, and those directions include guidance on how much users should rotate to turn at each waypoint.

These tools suffer from the same issues that tools within games have. They simply guide users to where they need to go, and do not offer users much agency when navigating, nor do they offer users a direct way of ``looking around'' environments themselves. Talking Points 3~\cite{Yang2011} features some degree of looking around, specifically a ``Directional Finder'' that provides a list of POIs in the general 45$\degree$-wide direction that a user points their phone. However, users cannot precisely look in a given direction and must physically turn their body toward a general direction to hear what lies that way. \replaced{Our preliminary exploration of NavStick~\cite{NairSmith2020} proposed the concept of surveying environments directionally; in this paper, we perform studies to evaluate NavStick with respect to existing navigation paradigms and investigate the role of such a system in video games.}{Nair and Smith~\cite{NairSmith2020} proposed a means of surveying environments directionally, but their work was only conceptual (without an empirical study), and it did not investigate the role of such a system in video games.}

Another issue with real-world navigation tools is that their automatic announcements may cause users to miss important environmental sounds~\cite{Dias2015} --- a common complaint among VIPs~\cite{Fallah2013}. NavStick only plays announcements or tones while the user is actively scrubbing.

% NavStick, by contrast, allows players to survey in any direction they want without needing to physically turn their character.

% Some tools, such as NavCog3 and Microsoft Soundscape, automatically announce landmarks and points of interest (POIs) that users pass as they navigate from place-to-place. Unlike NavStick, however, these systems' announcements are automatic and therefore provide neither \emph{on-demand} nor \emph{random} access to information about users' surroundings in all directions. Furthermore, the automatic announcements that these tools generate may cause users to miss important environmental sounds~\cite{Dias2015}, which is a common complaint among VIPs~\cite{Fallah2013}.

% Talking Points 3~\cite{Yang2011} offers VIPs a richer means of exploring the environment than NavCog3 and Microsoft Soundscape: it features a ``Directional Finder'' that provides a list of points of interest (POIs) in whichever direction a user is holding their phone. This feature, however, does not provide random access: the user can only survey the direction that they are facing at any one time. It also lacks precision: it outputs a list of all POIs within a $45\degree$ circular sector and within 100\,m, which makes pinpointing the precise direction of objects impossible. By contrast, NavStick allows users to survey in any direction they want without needing to physically turn their character in that direction \textit{and} communicates the exact relative direction of objects.

\section{NavStick}

NavStick is an audio navigation tool that gives VIPs the ability to \textit{look around} virtual environments in a self-directed way. NavStick was co-designed with VIPs and centers around the idea that navigation involves much more than simply being guided along a path efficiently or scrolling through a list of POIs --- that is, players should have on-demand, random access to information about their surroundings. Two major aspects of NavStick's design facilitate this kind of navigation: (1) the act of \textit{``scrubbing,''} in which the player can tilt a thumbstick circularly to ``look'' in different directions; and (2) a circular data structure that we call a \textit{``NavPie,''} which represents the player's immediate surroundings. 

\begin{figure}
    \centering
    \begin{subfigure}{0.45\columnwidth}
        \includegraphics[width = \columnwidth]{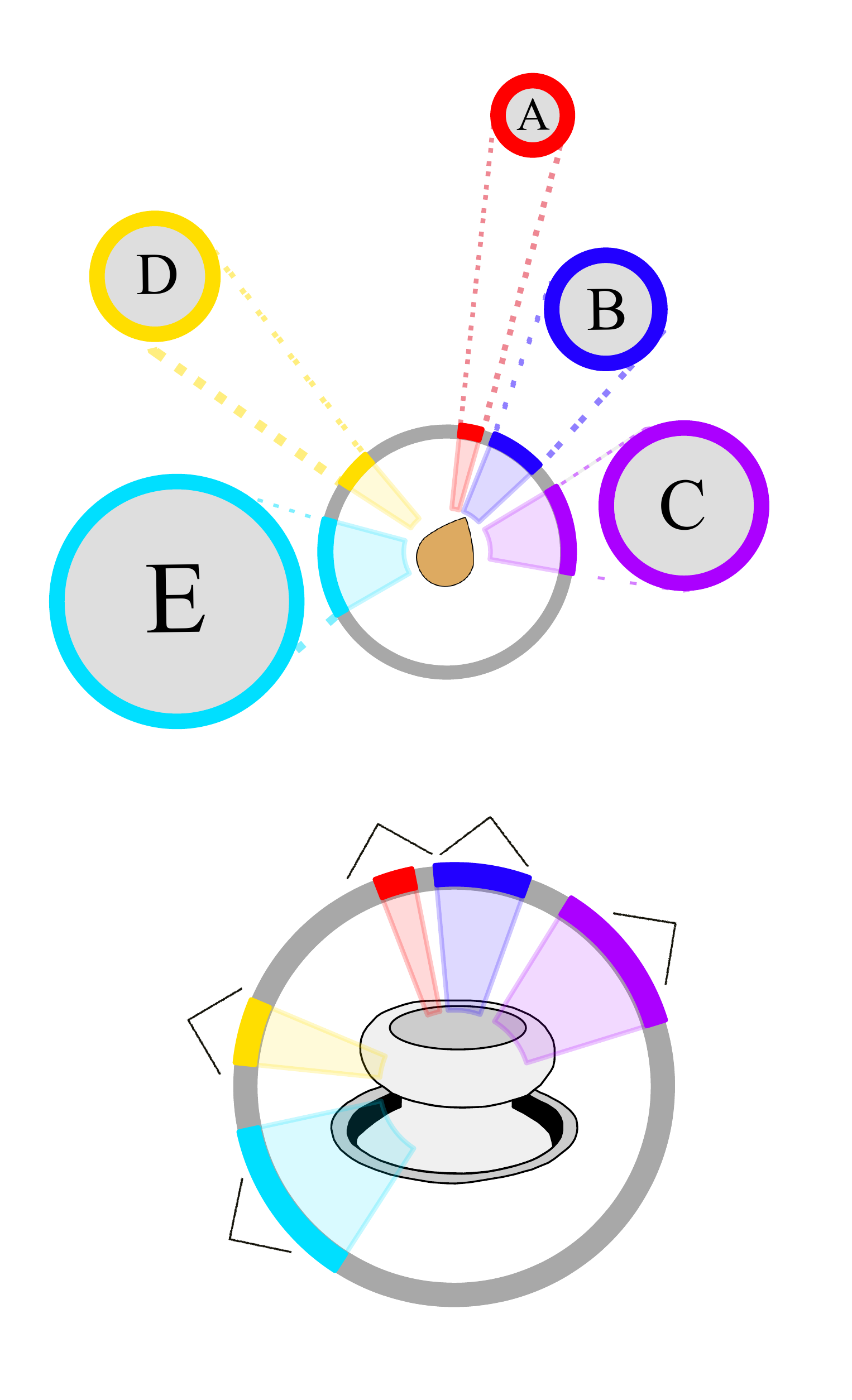}
        \caption{}
        %\label{fig:accy_over_time_prev}
        %\vspace{12pt}
    \end{subfigure}
    \hspace*{0.4em} % separation between the subfigures
    \begin{subfigure}{0.45\columnwidth}
        \includegraphics[width = \columnwidth]{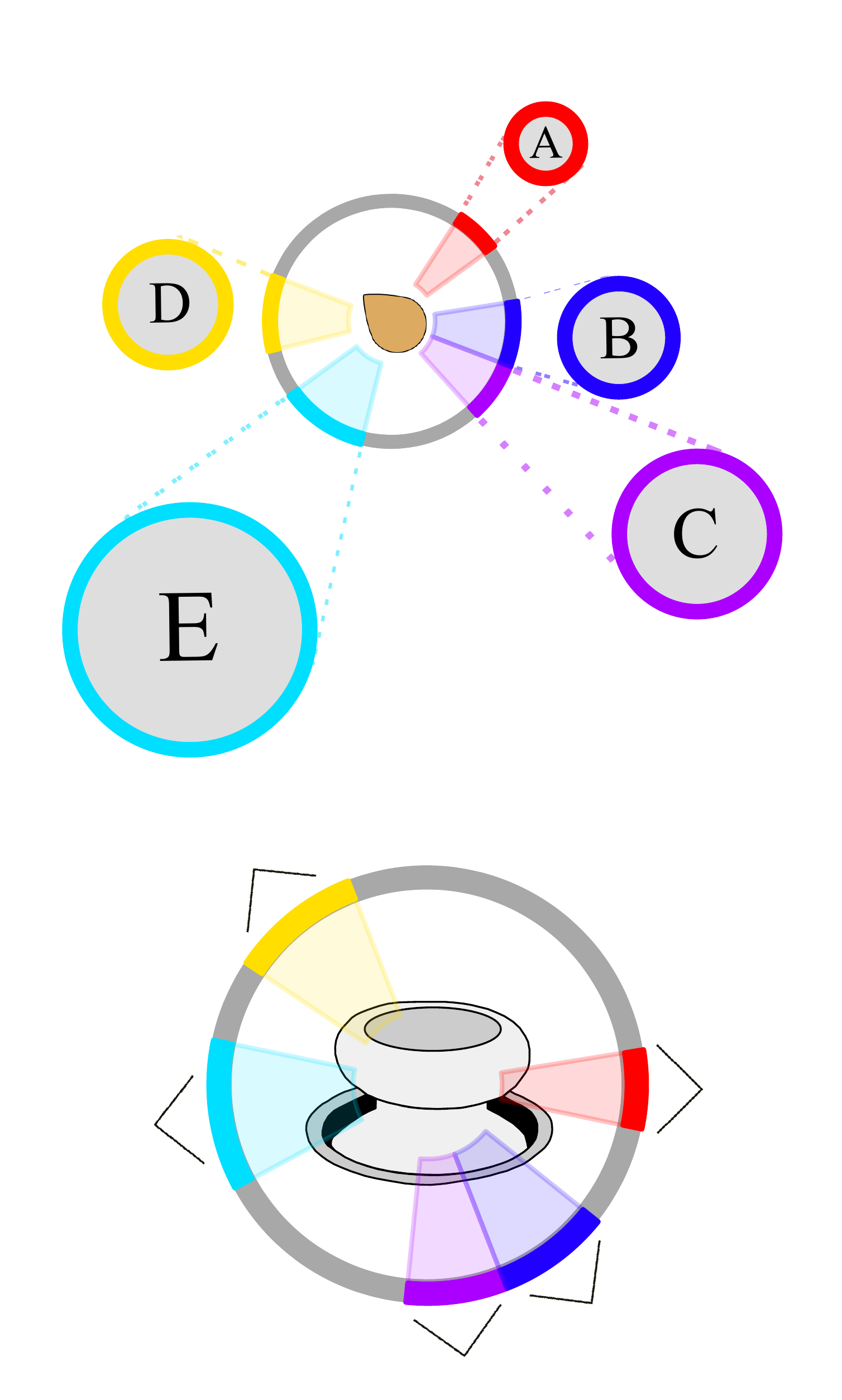}
        \caption{}
        %\label{fig:accy_over_time_new}
    \end{subfigure}
\caption{Representing a player's surroundings with NavPies. We show a player surrounded by five objects (A--E) of varying sizes. \textit{Top:} Two different player positions among these objects. \textit{Bottom:} Forward perspective NavPies. Here, each pie has been rotated such that it visualizes the pie from the player's forward perspective. The arrow outside each colored slice represents the direction that the thumbstick must be tilted in order to ``look at'' each object.}
\label{fig:navpie}
\end{figure}

\subsection{Scrubbing}

NavStick repurposes a game controller's right thumbstick into a tool for looking around. To use it, players tilt the thumbstick in any direction to hear (via speech) what lies in that precise direction via a raycast. They can then ``scrub'' the thumbstick circularly around the edge to quickly survey many different directions. Figure~\ref{fig:navstick_overview} illustrates how to use NavStick. In the figure, the player is tilting the right stick in the one o'clock direction. NavStick announces what lies in that direction from the player's current perspective --- in this case, Chomper \#4 (an enemy) --- using spatialized audio emanating from the selected item itself. \added{In the current iteration of NavStick, raycasts are performed straight out from the eye-level of the in-game character that the player controls --- that is, parallel to the ground. (We discuss controlling NavStick's pitch angle as a point of future work in Section \ref{sec:vertical-surveying}.)}

NavStick operates via line-of-sight \added{(like a laser pointer)}, only taking into account what is \added{directly} visible from the player's \replaced{current position}{point of view}. \added{NavStick will only announce the first object situated in a given direction and has an infinite scan radius.} We map NavStick to the right thumbstick because most 3D games use the right thumbstick to control the in-game camera. Controlling the game's camera is not useful for VIPs since many VIPs lack vision, so NavStick repurposes this thumbstick to act as a navigation aid instead. \added{Similar to how a sighted player can move and control the camera at the same time, VIPs can also move and use NavStick simultaneously.}

NavStick's design is inspired by earPod~\cite{Zhao2007}, which repurposed an early iPod's click wheel into a tool for traversing eyes-free menus such as telephone menus. Similar to earPod, any speech or tone played by NavStick at a slice is immediately truncated when the user moves off the slice\added{. Even if the user scrubs quickly, they will still be able to hear the first syllable of an announcement}, providing users with a strong feeling of reactivity.

We implemented NavStick in the Unity game engine (version 2020.1.16f1)~\cite{UnityTechnologies2020}. Players interact with the world using a standard game controller and receive feedback via audio using a standard pair of headphones. To generate spatialized sound, we used the Steam Audio toolkit for Unity~\cite{ValveSoftware2019}, which provides spatialized audio via its built-in head-related transfer function (HRTF). We pre-synthesized speech output using the Google Cloud Text-to-Speech service~\cite{GoogleCloudTTS}.

\subsection{NavPies}

To represent a player's surroundings, NavStick uses a circular data structure called a NavPie. NavPies can be thought of as spatial versions of pie menus; each slice represents where an object is in relation to the player at that moment in time. Each of the NavPie's ``slices'' or ``regions'' represents an item that is visible from the player's point of view. The slice's direction and circular extent (i.e., arc length) represent the object's direction and angular extent (the angle that the object subtends with respect to the player's location).

Figure~\ref{fig:navpie} shows how NavStick computes NavPies from a player's surroundings. The figure depicts a player at two different positions (top-left and top-right) among five objects (labeled A--E). The circle surrounding the player is the NavPie, and its arcs (``slices'') are formed via line-of-sight ray traces. Each arc represents an object. Note that the arcs do not cover the NavPies' full circumferences. ``Empty'' slices (gray arcs) indicate that there is nothing of interest in that direction in the player's immediate surroundings. \replaced{NavStick can be configured to either play no audio at all (as in Study 1) or play a default tone (as in Study 2) when players tilt or scrub in these directions.}{NavStick can be configured to either play a default tone (as in Study 1) or not play audio at all (as in Study 2) when players tilt or scrub in these directions.}

The NavPies on the bottom-left and bottom-right of Figure~\ref{fig:navpie} are the same ones as shown in the top-left and top-right (respectively) but from the player's forward perspective, where 12 o'clock indicates the player's forward direction. The arrows around these NavPies show how NavStick maps the NavPies' slices to thumbstick directions.

\subsection{Supported Navigation Tasks}

NavStick supports all three navigation tasks that Darken and Sibert~\cite{Darken_Sibert_1996} classify for virtual worlds: naïve search, primed search, and exploration. In addition, NavStick supports both novice and expert usage patterns. Novices, who are unfamiliar with their surroundings, might choose to scrub NavStick's entire circumference to survey their full surroundings, thereby performing an exhaustive search (\textit{naïve search}). Experts, who are already somewhat familiar with their surroundings, can choose to scrub in specific directions of interest to perform a non-exhaustive search (\textit{primed search}). Both groups of users can also use NavStick to form a cognitive map even when there is no target (\textit{exploration}).

\begin{figure}[]
  \centering
  \includegraphics[width=0.95\linewidth]{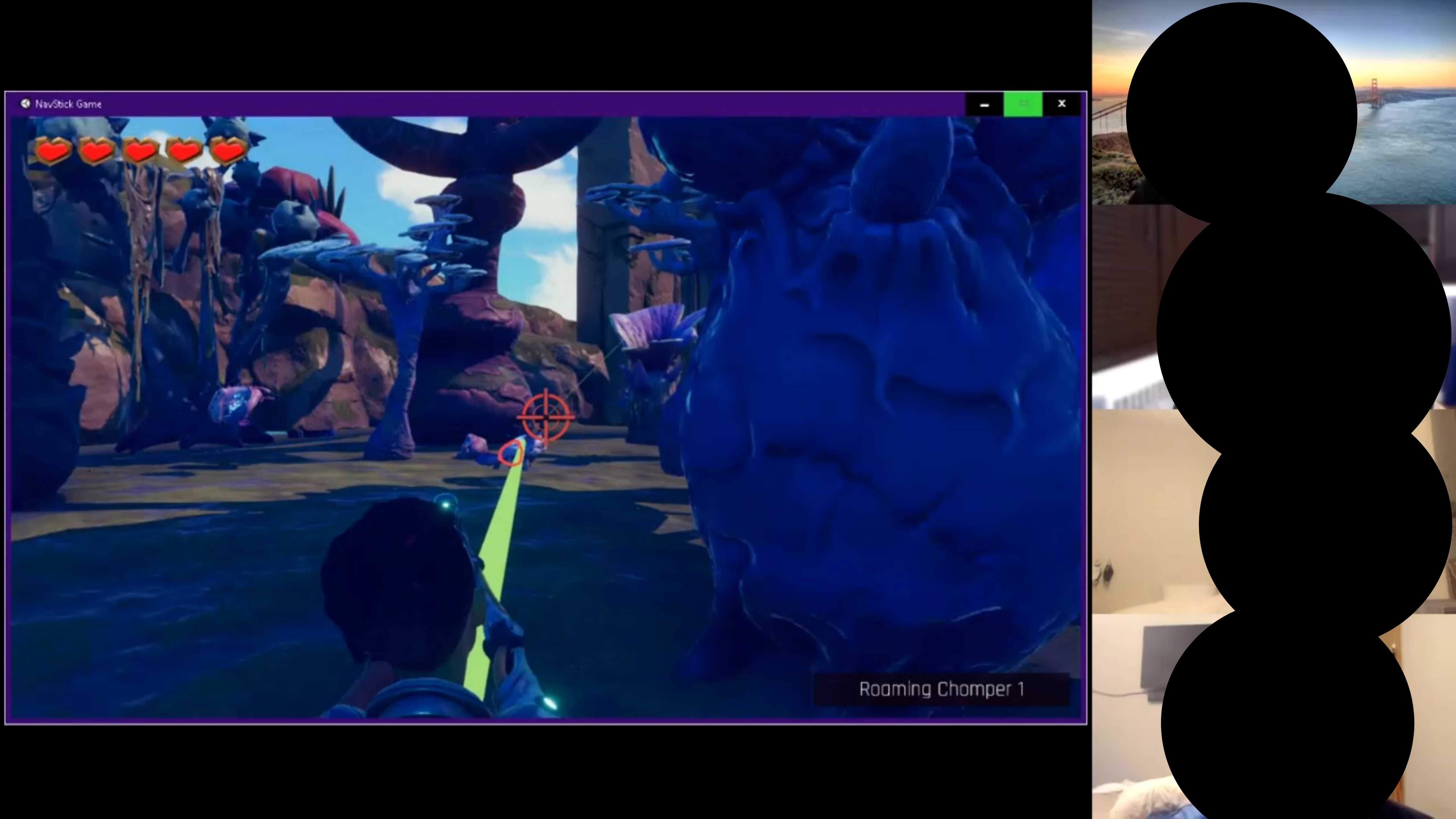}
  \caption{Remote study session with a participant for Study \replaced{2}{1}. Here, a study participant is currently traversing Segment 8 of our prototype 3D adventure game, \textit{The Explorer}. \textit{(Faces obscured to protect anonymity.)}}
  \label{fig:zoom_session}
\end{figure}

\begin{table*}[ht]
\begin{tabular}{c|ccc}
\#         & \textbf{Navigation Task}      & \textbf{Repetitions}            & \textbf{Goal}                                   \\ \hline
\textbf{1} & Exploration \textit{(directional)}     & (2 aisles, 4 Qs each) x 2 tools & Look first, then hear Q about item \textit{locations}. \\
\textbf{2} & Exploration \textit{(non-directional)} & (3 aisles, 1 Q each) x 2 tools  & Look first, then hear Q about item \textit{presence}. \\
\textbf{3} & Naïve search \textit{(directional)}          & (3 aisles, 1 Q each) x 2 tools  & Hear Q about item \textit{locations}, then look.       \\
\textbf{4} & Naïve search \textit{(non-directional)}      & (3 aisles, 1 Q each) x 2 tools  & Hear Q about item \textit{presence}, then look.       \\
\textbf{5} & Primed search                    & (3 aisles, 1 Q each) x 2 tools  & Compare prices of items of the same type.       \\
\textbf{6} & Video game navigation                    & (1 level) x 2 tools             & Traverse a game level (room from \textit{Terraformers}).
\end{tabular}
\vspace{2mm}
\caption{An overview of the six navigation tasks for Study \replaced{1}{2}. \added{(Q = question.)}}
\label{table:task_summary}
\end{table*}

\section{COVID-19 Challenges}

The COVID-19 pandemic prevented us from evaluating NavStick in-person, so we were faced with the task of designing a completely remote study. By its very nature, however, NavStick presents distinct advantages for a remote study: the only hardware required is a game controller and a standard pair of headphones. Our prototype was implemented in Unity, which can compile executables for all major computing platforms. Executing the study remotely also allowed us to reach VIPs who may not feel comfortable traveling too far from home or who may be otherwise isolated, due to challenges in navigating the real world.

To execute our two studies in an entirely remote manner, we packaged the Unity study environment into portable game executables for both Windows and macOS. Each participant had their own executable that was signed with a unique identifier. We sent each participant a link to download their executable before their study appointment. We held the study appointments over Zoom and asked participants to share their computer audio (and, optionally, video) with the facilitators. Figure~\ref{fig:zoom_session} shows a Study \replaced{2}{1} session in progress. Both studies were approved by \replaced{the Columbia University}{our university's} Institutional Review Board (IRB).

We designed the executables to connect with a cloud backend that synchronizes with the runtime state of participants' games. This allowed us to control the executables remotely as needed. We could, for example, enable or disable rendered graphics, advance to the next study condition/segment, repeat a trial, and tweak other options fully remotely using a custom-built control panel.

%%%

\section{Study 1: NavStick vs. Linear Menus for Different Navigation Tasks}

\comment{Studies 1 and 2 were swapped. The comparison study is now S1; the game study is now S2.}

\deleted{In Study 1, we analyzed NavStick's performance across various game \textit{environments}, revealing its strengths and limitations in representative video game settings.} In Study \replaced{1}{2},\replaced{ we compare NavStick with the ``status quo'' means of surveying virtual environments: using linear menus that list surrounding items.}{ we complement Study 1 with a full study on how NavStick performs across a broad range of navigation \textit{tasks}.} Prior work in spatial cognition~\cite{Golledge_1999} and navigation tools~\cite{Darken_Sibert_1996} show that navigation tools should be analyzed and evaluated \deleted{not only with respect to various environments but also }with respect to navigation tasks such as exploration vs.\ search. \replaced{As such, w}{W}e will also use Study \replaced{1}{2} as an opportunity to \replaced{assess how NavStick performs across a broad range of navigation \textit{tasks}}{compare NavStick with the ``status quo'' means of surveying virtual environments: using linear menus that list surrounding items}.

To represent status quo surveying, we implemented ``NavMenu,'' a linear menu that alphabetically lists POIs in the player's current area. Players ``open'' the NavMenu by pressing the left shoulder button, then navigate up and down within the menu by using the D-pad. When the player opens the menu, the system announces the number of targets in the area. Each time the player presses up or down on the D-pad, they will hear the corresponding item announced via speech spatially rendered at the item's location. Recall that NavStick also announces items via spatialized speech, but does so when the player tilts the thumbstick toward an item.

\added{Although we opted for an alphabetically-sorted NavMenu for this study, we also initially considered a proximity-ordered menu (where elements are sorted based on how far they are from the player) and a directionally-ordered menu (where elements are sorted based on their direction relative to the player). Consecutive elements in both alphabetical- and proximity-ordered menus can be located in scattered directions. However, proximity-ordered menus have an additional weakness where the menu's ordering will change as the player moves through the world. Traversing a directionally-ordered menu is equivalent to scrubbing continuously around with NavStick, but with more effort and without the on-demand, random access to items that NavStick affords. As a result, an alphabetically-sorted NavMenu provides the strongest possible comparison with NavStick.}

We investigate the following research questions in Study \replaced{1}{2}:
%\vspace{-5mm}
\begin{enumerate}[label=(RQ\replaced{1}{2}.\arabic*), leftmargin=5.1\parindent]
    \item What are NavStick's advantages and disadvantages compared to menu-based surveying (NavMenu)?
    \item Which forms of navigation is NavStick better suited for, and which forms is NavMenu better suited for?
\end{enumerate}

\subsection{Study 1 Navigation Tasks}

We asked participants to perform six types of navigation tasks using both NavStick and NavMenu. Table \ref{table:task_summary} shows an overview of these tasks. Tasks 1--5 involved navigating within an aisle in a virtual grocery store, and Task 6 involved completing a level from the video game \textit{Terraformers}~\cite{PinInteractive2003a, Westin2004}. These tasks, although highly-controlled, allowed us to compare NavStick and NavMenu for different virtual navigation purposes.

\begin{figure}[]
  \centering
  \includegraphics[width=0.95\linewidth]{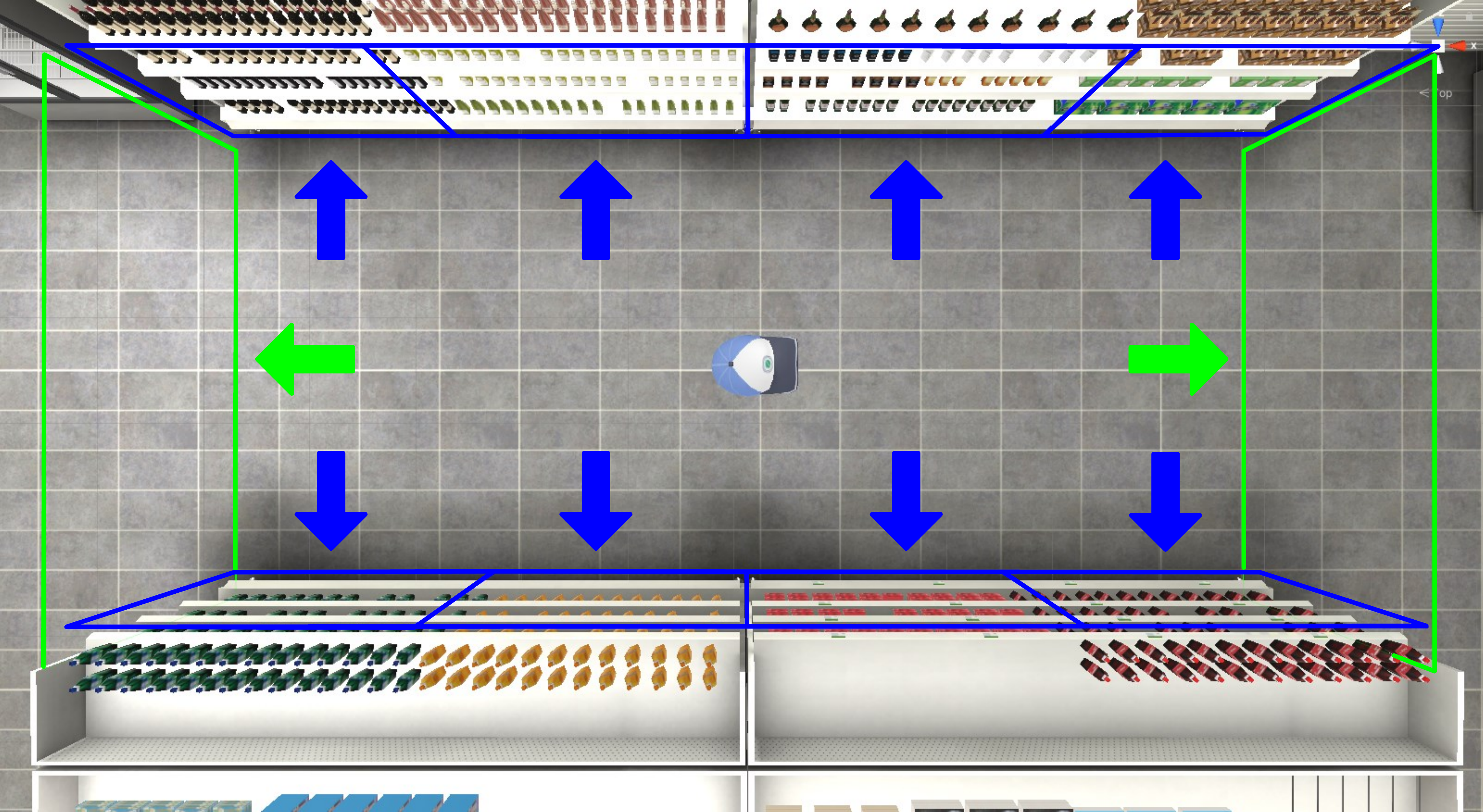}
  \caption{Overhead view of our virtual grocery store aisle for Study \replaced{1}{2}. The dark blue arrows point to the 8 targets in the aisle. Each target is represented by a 2D rectangle outlined in dark blue and positioned right in front of the shelf. The light green arrows point to 2 invisible walls (also in light green) that prevent players from moving outside the aisle.}
  \label{fig:aisle}
\end{figure}

\subsubsection{Grocery store navigation tasks \textit{\textbf{[Tasks 1--5]}}}

\hfill\newline We asked participants to perform tasks that spanned Darken and Sibert's classification of wayfinding tasks for virtual worlds~\cite{Darken_Sibert_1996}. \textbf{Exploration} tasks do not involve a target in the initial search and require forming a cognitive map for later use. In \textbf{naïve search} tasks, the player has a target but does not know its location, possibly requiring an exhaustive search of the NavPie in the worst case. In \textbf{primed search} tasks, the player knows that the target items are laid out in front of them, and so does not have to scrub the entire NavPie to find the items.

We chose a virtual grocery store for this study since it represents a place in which people routinely perform all of these different tasks. Figure \ref{fig:aisle} shows the grocery store aisle layout that we used. It features eight equal-length item ``slots'' (four on each side) that we placed items into to represent different aisles such as canned goods or beverage aisles. \added{These targets were made large enough such that that they were ''airtight'' (i.e., they had no gaps between them).}

For the \textbf{exploration} tasks (Tasks 1 \& 2), we first let participants explore and survey the aisle for as long as they wished, then we asked them one of two types of questions. For Task 1, we asked them to name an item that is located close to a randomly chosen item. The closer their answer was, the better. For Task 2, we gave the participant a ``shopping list'' of five items, and they had to say whether or not each of the items was present in the aisle.

For the \textbf{naïve search} tasks (Tasks 3 \& 4), we asked participants the same questions as Tasks 1 \& 2, respectively. The difference is that participants surveyed the aisle \textit{after} hearing the question.

We refer to Tasks 1 \& 3 as ``directional'' in Table~\ref{table:task_summary} because they test participants' knowledge of how items are \textit{directionally} situated with respect to one another. Similarly, we refer to Tasks 2 \& 4 as ``non-directional'' in Table~\ref{table:task_summary} because participants do not need to remember items' directions --- only whether or not they are present.

In the \textbf{primed search} task (Task 5), we positioned participants in front of a group of items and asked them to find the cheapest one among that group. Players can hear an item’s price by pressing the right bumper \textit{(RB)} while selecting it with NavStick or NavMenu. 

\added{For all tasks, we informed participants that the aisle contained eight items, and all participants surveyed all items in the aisle before committing to an answer (for naïve search and primed search) or hearing the question (for exploration). }

\begin{figure}[]
  \centering
  \includegraphics[width=0.95\linewidth]{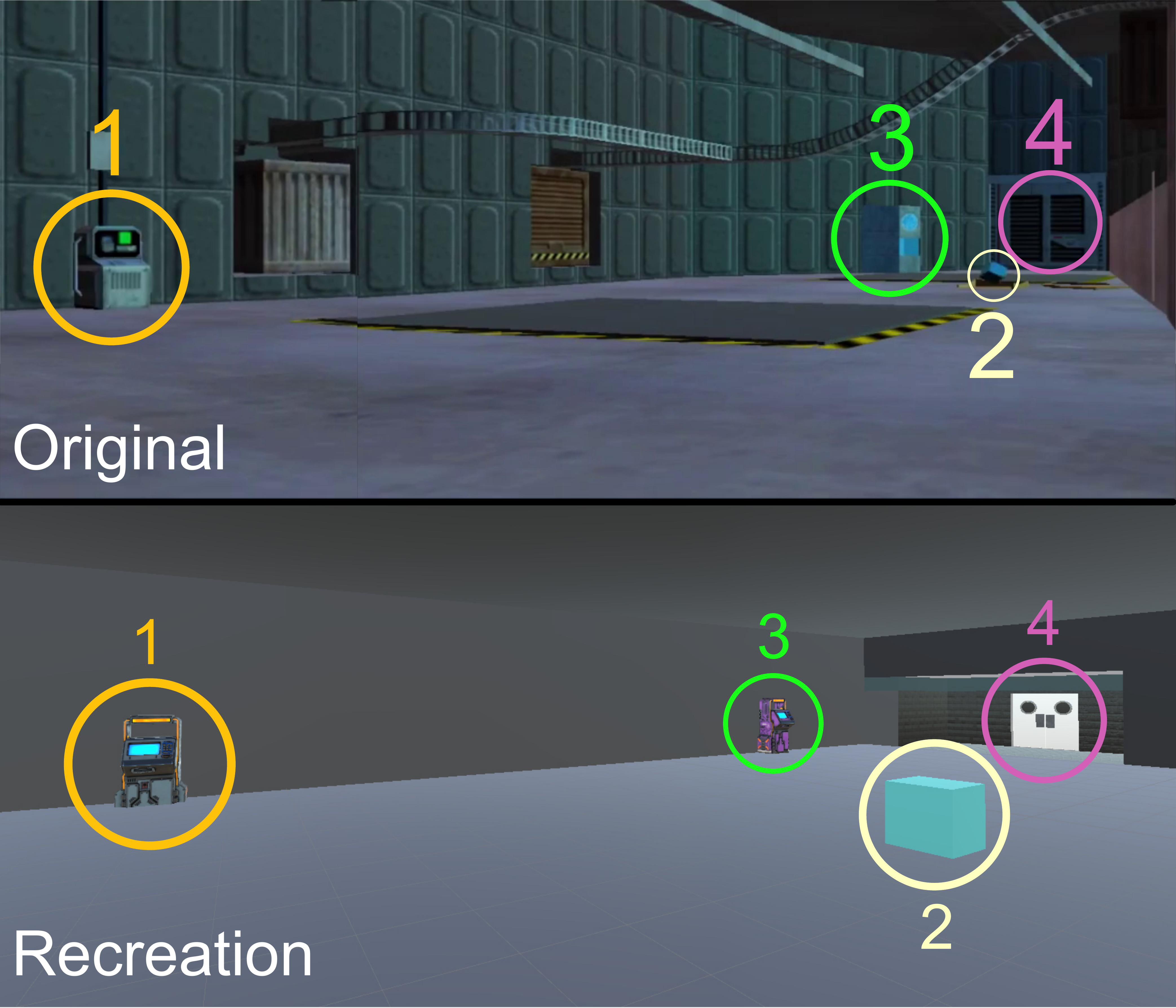}
  \caption{Comparison of \textit{Terraformers} environments. \textit{Top:} The room as shown in the original game~\cite{PinInteractive2003a, Westin2004}. \textit{Bottom:} The room as replicated by us in Unity. The room consists of four objects: a computer terminal (\#1) that reveals the location of a key (\#2) which must be inserted into a keyhole (\#3) in order to open an exit door (\#4).}
  \label{fig:terraformers}
\end{figure}

\begin{figure}[t]
    \centering
    \includegraphics[width=0.95\linewidth]{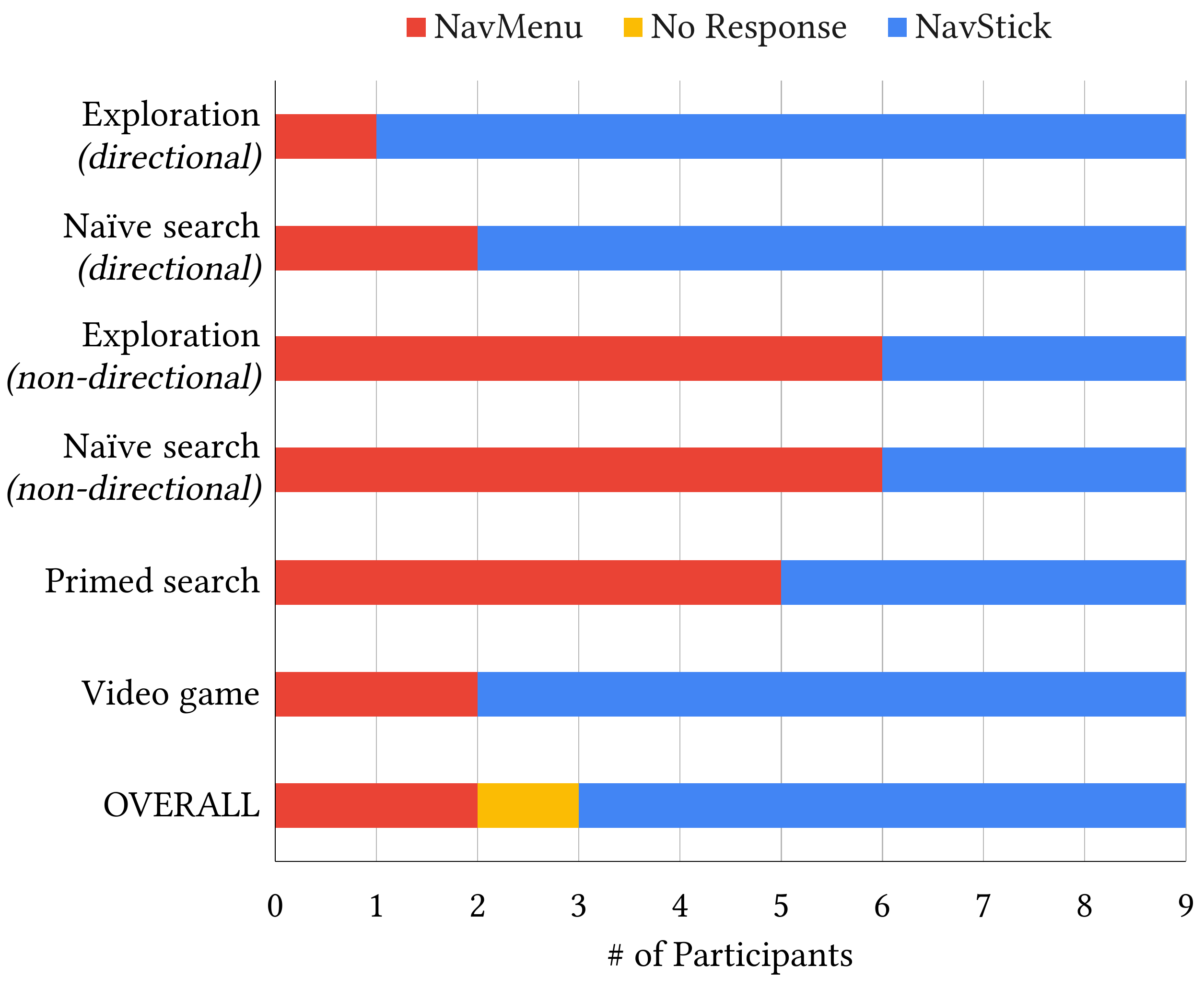}
    \caption{Preferred navigation tool for each type of navigation task. Most participants preferred NavStick for directional tasks (in which the location of items is important) and for the video game task. Most preferred NavMenu, however, for non-directional tasks: ones in which it is only important to understand \textit{what} items are present. Overall, participants preferred NavStick if forced to choose, but strongly felt the two interfaces worked hand in hand. One refused to respond for this reason.}
    \label{fig:pref_chart}
\end{figure}

\subsubsection{Video game navigation task \textbf{[Task 6]}.}

\hfill\newline Task 6 is a video game level that we recreated from \textit{Terraformers}, a blind-accessible video game~\cite{PinInteractive2003a, Westin2004}. Figure \ref{fig:terraformers} illustrates the level. \replaced{As NavStick's goal is to make 3D \textit{games} more blind-accessible}{Since we did not compare NavStick with any other navigation tool in Study 1}, we wanted to use Task 6 to compare NavStick and NavMenu within a game context. We chose \textit{Terraformers} because\deleted{many of our Study 1 participants mentioned} it is a classic audio game \replaced{that features}{with} linear menus for navigating.

To complete the level, players must navigate to four targets in sequence within the room: a computer terminal, a key, a keyhole, and an exit door. Unlike our virtual grocery store (Tasks 1--5), this game level includes ambient environmental sounds similar to the original sounds from \textit{Terraformers}. When players focus on a target using NavStick or NavMenu, they can press the left bumper \textit{(LB)} on their controller to teleport to that target’s location instantly.

\subsection{Study 1 Participants}

% We recruited nine participants for Study \replaced{1}{2}, all of whom described themselves as male and having no usable vision. \deleted{Five had participated in Study 1. }Five participants (P2-4, P7, P9) were 18-25 years old and the other four (P1, P5, P6, P8) were 26-35 years old. P6 reported minor hearing loss in their right ear. All but two participants (P7 and P8) reported themselves as experienced with video games (4+ on a 5-point Likert scale). We recruited these participants\replaced{ through posts on the AudioGames.net Forum,\footnote{\url{https://forum.audiogames.net/}} an online discussion board that centers around audio-based games and is frequented by VIPs.}{ in the same way as Study 1.}

We recruited nine participants for Study \replaced{1}{2}, all of whom described themselves as male and having no usable vision. \deleted{Five had participated in Study 1. }Five participants (P2-4, P7, P9) were 18-25 years old and the other four (P1, P5, P6, P8) were 26-35 years old. P6 reported minor hearing loss in their right ear. All but two participants (P7 and P8) reported themselves as experienced with video games (4+ on a 5-point Likert scale). We recruited these participants through posts on the AudioGames.net Forum,\footnote{\url{https://forum.audiogames.net/}} an online discussion board that centers around audio-based games and is frequented by VIPs.

\subsection{Study 1 Procedure}

We began by giving participants unlimited time to experiment with NavStick and NavMenu within a trial aisle. Afterward, each participant performed Tasks 1--6 in that order, but they were assigned a counterbalanced order for using NavStick and NavMenu for the tasks. After every (tool, task) combination, we administered a post-task questionnaire that gauged participants' perceived effort, frustration, and impressions of using the assigned tool for the task. The questionnaire employed 20-point Likert scales similar to a NASA TLX form~\cite{Hart1988}.

For exploration tasks (Tasks 1--2), we also asked participants how well they thought they were able to create a mental map of their surroundings. For the video game task (Task 6), we also asked how fun it was to play the level with the tool they used. All responses were given on a 20-point Likert scale. After performing each task with both NavStick and NavMenu, we asked participants to state which tool they preferred for the task.

At the very end, we administered a post-study questionnaire followed by a semi-structured interview to ask participants to reflect on what they liked and disliked about both tools and to choose their favorite in a forced ranking.

\begin{figure}[t]
    \centering
    \includegraphics[width=0.95\linewidth]{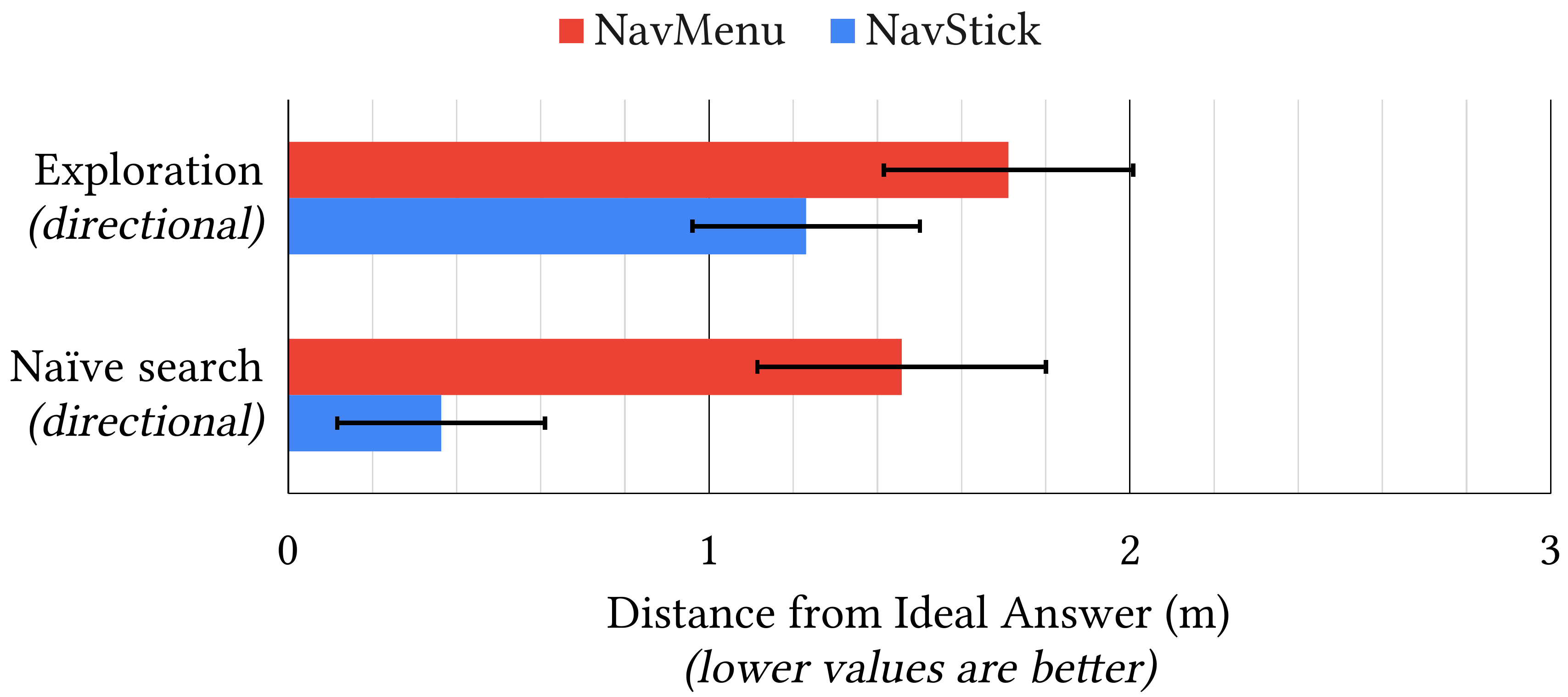}
    \caption{Average distance from ideal answers for \textit{directional} navigation tasks (Tasks 1 \& 3) using NavMenu and NavStick. An \textit{ideal answer} is an item that is adjacent to the item named in the prompt and has a distance of 0. This figure objectively measures the ``accuracy'' of users' mental maps after using NavMenu and NavStick. Lower numbers are better; error bars indicate standard error.}
    \label{fig:dir_accuracy_chart}
\end{figure}

\begin{figure}[]
    \centering
    \includegraphics[width=0.95\linewidth]{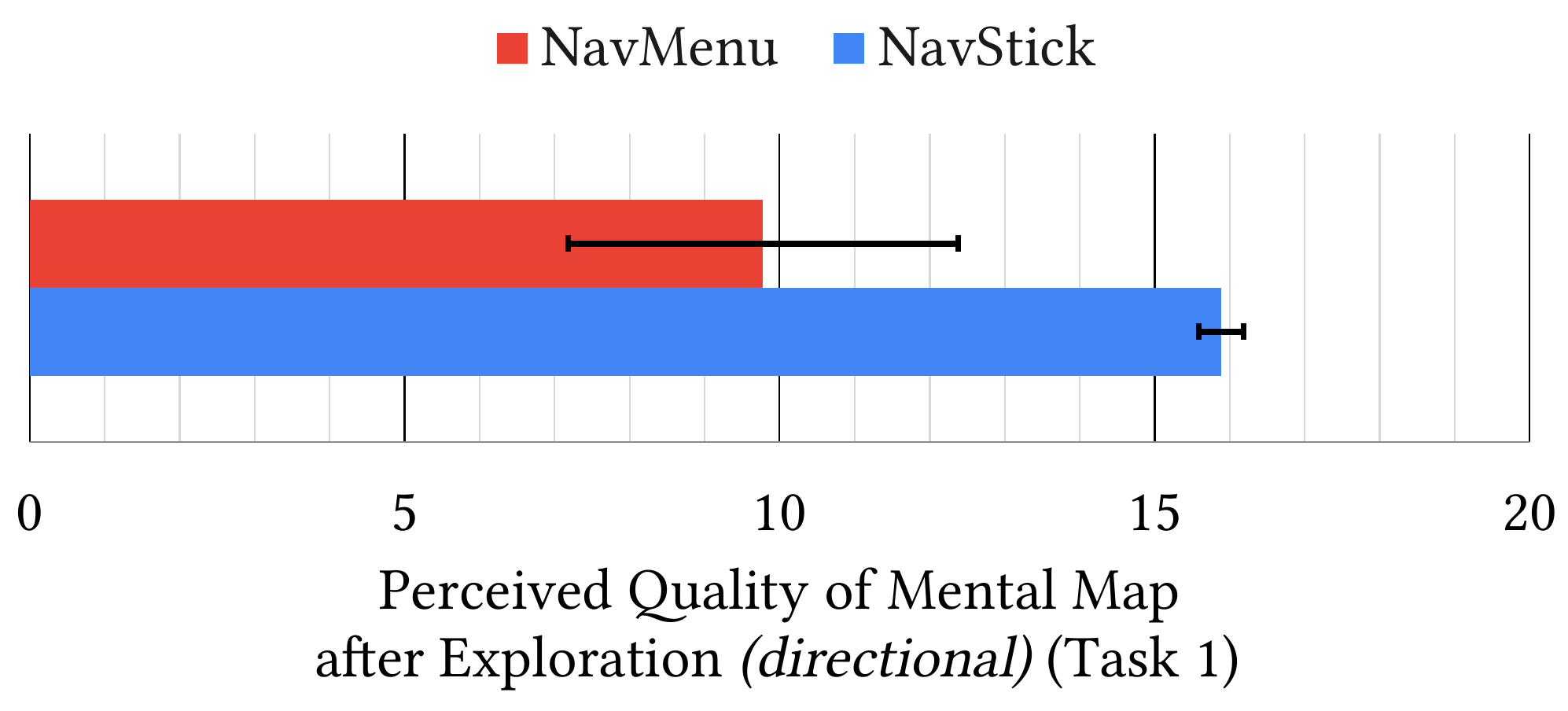}
    \caption{Average perceived quality of mental map for each navigation tool after the directional exploration task (Task 1). Error bars indicate standard error.}
    \label{fig:dir_mental_map_chart}
\end{figure}

\subsection{Study 1 Results}

Below, we report the main takeaways from Study \replaced{1}{2} in terms of answering our research questions: learning what NavStick's advantages and disadvantages are compared to NavMenu (RQ\replaced{1}{2}.1), and learning which forms of navigation each tool is better suited for (RQ\replaced{1}{2}.2). We analyzed our qualitative results (participant quotes\deleted{ and Likert scale results}) \replaced{via qualitative coding. All authors collectively clustered data into common themes, debating among themselves until there was mutual agreement on the themes. Each theme presented below represents opinions shared by a majority of our participants}{identically to Study 1}. \deleted{Below, we share our findings on the advantages NavStick has over NavMenu as well as the advantages NavMenu has over NavStick.} For statistical significance conclusions \added{with respect to our quantitative results (Likert scale responses and in-game data)}, we ran Welch’s t-tests on the data to obtain two-tailed p-values and assumed a 95\% confidence interval. \added{Below, we share our findings on the advantages and disadvantages that each of the two tools --- NavStick and NavMenu --- have over the other.}

\begin{figure}[t]
    \centering
    \includegraphics[width=0.95\linewidth]{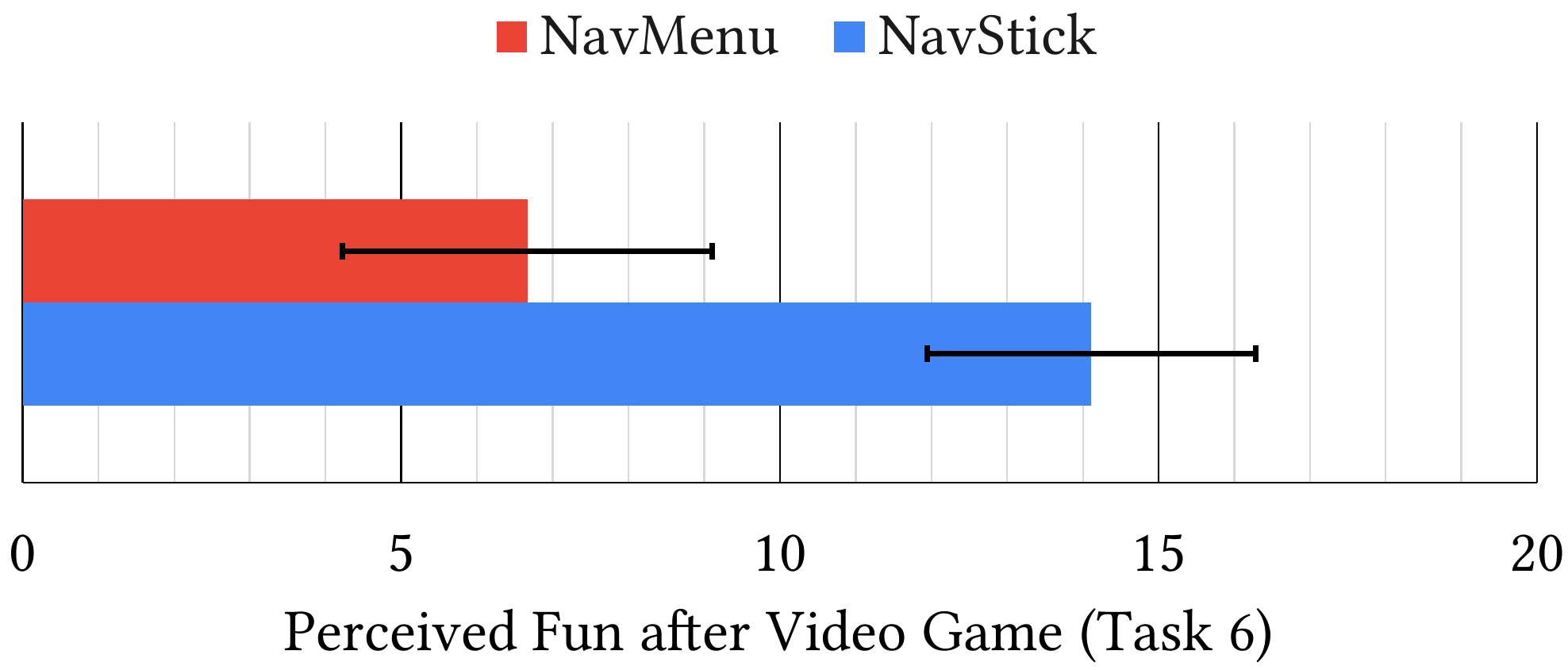}
    \caption{Average perceived fun for each navigation tool for the video game task (Task 6). Error bars indicate standard error.}
    \label{fig:tf_fun_chart}
\end{figure}

\begin{figure}[]
    \centering
    \includegraphics[width=0.95\linewidth]{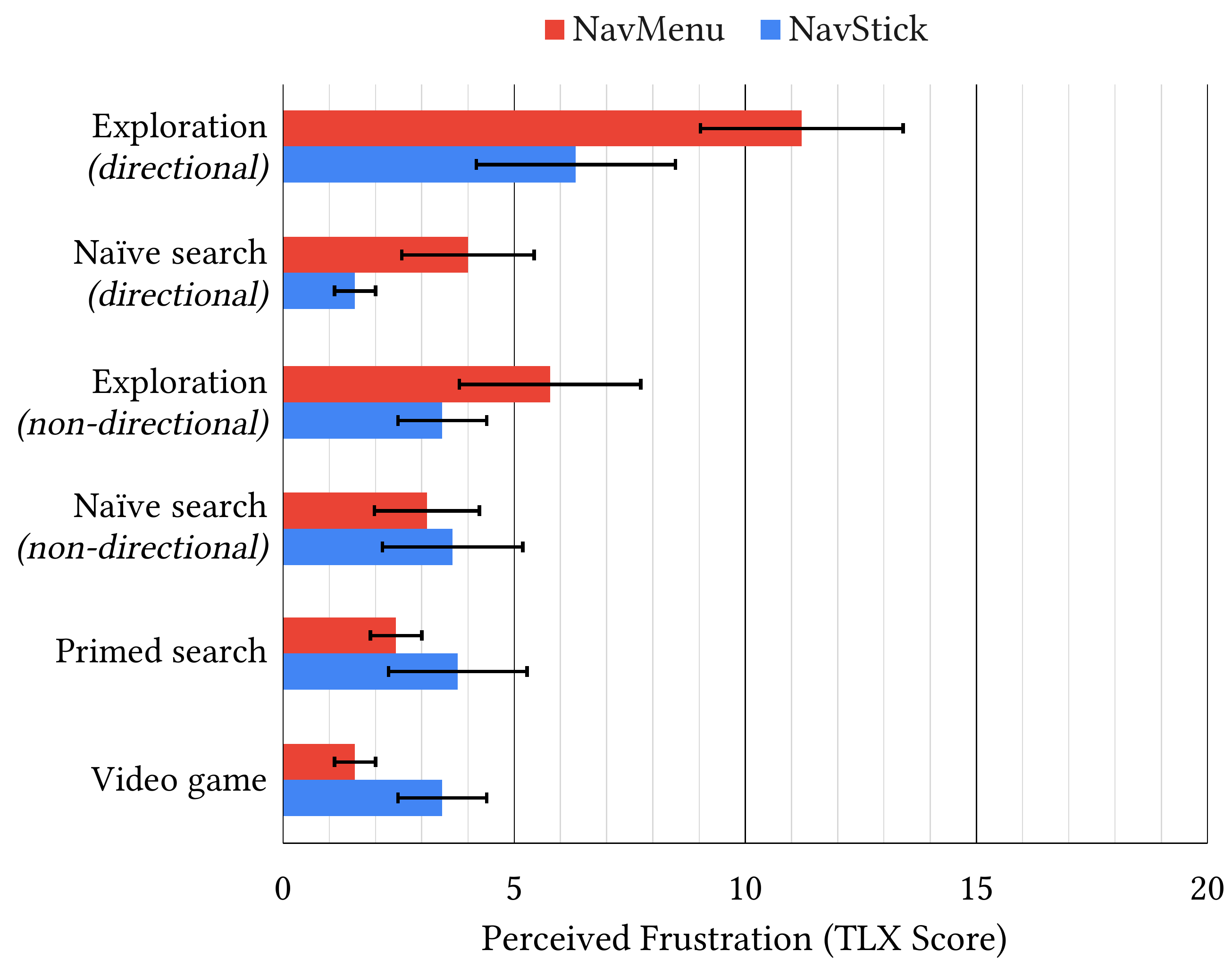}
    \caption{Average perceived frustration for each navigation tool per navigation task. Error bars indicate standard error.}
    \label{fig:frustration_chart}
\end{figure}

\subsubsection{Results: NavStick's Advantages over NavMenu}

\hfill\newline As Figure~\ref{fig:pref_chart} shows, participants preferred NavStick for directional tasks (Tasks 1 \& 3) and the video game task (Task 6).

Figure \ref{fig:dir_accuracy_chart} shows participants' average distance from ideal answers for \textit{directional} navigation tasks (Tasks 1 \& 3) using NavMenu and NavStick. An \textit{ideal answer} is an item that is adjacent to the item named in the prompt (``Name an item that is close to [item X].''). The distances are measured in meters. They act as an objective measure of the ``accuracy'' of users' mental maps after using NavMenu and NavStick.

For directional naïve search (Task 3), we found a \textit{significant} main effect of navigation tool (NavMenu vs.\ NavStick); answers given using NavStick ($M=2.7$, $SD=0.7$) were closer to the item in the prompt than ones given using NavMenu ($M=3.8$, $SD=1.0$) ($p < 0.05$).

Figure~\ref{fig:dir_mental_map_chart} shows participants' perceived quality of their mental maps after exploring with NavMenu and NavStick. Participants felt that they were able to form \textit{significantly} better mental maps with NavStick ($M=14.5$, $SD=5.2$) than with NavMenu ($M=9.9$, $SD=7.8$) ($p < 0.05$). This finding is consistent with our objective finding in Figure \ref{fig:dir_accuracy_chart}. 

Figure \ref{fig:tf_fun_chart} shows the amount of fun participants reported having while playing the \textit{Terraformers} level (Task 6) using NavMenu and NavStick. Participants saw NavStick ($M=14.1$, $SD=6.5$) as \textit{significantly} more fun to use than NavMenu ($M=6.7$, $SD=7.3$) ($p < 0.05$). 

Since NavMenu's list-based format does not communicate information about directionality, participants struggled to understand where items were positioned relative to one another using NavMenu. Figure \ref{fig:frustration_chart} shows how participants rated NavMenu and NavStick in terms of frustration. We found that NavStick ($M=5.8$, $SD=3.9$) was significantly less frustrating to use than NavMenu ($M=7.6$, $SD=6.5$) for directional navigation tasks ($p < 0.05$). Verbal sentiments indicated that some players found NavMenu to be mentally demanding and confusing:

\begin{quote}
    \textit{"I’m so lost! I think some items are close to each other, but I keep scrolling and so it’s confusing."}  - \textbf{P3}
\end{quote}

Indeed, participants often had to take multiple passes through the NavMenu in order to keep track of the items present. We originally expected NavMenu to be significantly less frustrating and effortful than NavStick for the ``non-directional'' tasks (Tasks 2 \& 4) given that it specializes in listing out what items are present in a space. However, we did not find any such relationship between NavStick and NavMenu in either effort (NS: $M=5.2$, $SD=3.9$; NM: $M=5.3$, $SD=4.8$; $p > 0.05$) or frustration (NS: $M=3.6$, $SD=3.7$; NM: $M=4.4$, $SD=4.9$; $p > 0.05$).

\subsubsection{Results: NavMenu’s Advantages over NavStick}

\hfill\newline Although NavMenu's list-based format strips away information about the environment's layout, doing so has a benefit: it makes NavMenu quite effective at indicating what objects are present in the environment. Many participants noted that NavMenu's simple presentation allowed them to easily and completely understand the area's content:

\begin{quote}
    \textit{"I wasn’t even trying to mental map it. [...] I just wanted to know if it was there or not. So, I wasn’t worrying about where it actually was. I didn’t even care."} - \textbf{P6}
\end{quote}

Indeed, Figure \ref{fig:pref_chart} shows that, when forced to choose between using NavStick and NavMenu for the ``non-directional'' tasks (Tasks 2 \& 4) --- ones in which it was only important to understand \textit{what} was in the aisle and not \textit{where} they were located --- participants mostly chose NavMenu.

Furthermore, NavMenu presented POIs in discrete units accessible one-by-one with the D-pad. With NavStick, participants had to contend with small NavPie slices from items that were further away\deleted{, which made it was easy for them to skip over small slices when scrubbing around}. Several participants expressed their preference for the NavMenu for this reason, revealing a need to somehow optimize NavStick to handle thin slices better:

\begin{quote}
    \textit{“What I liked about [the NavMenu] is the use of the D-pad. It made things easier to navigate with instead of using [NavStick] in a way because with [NavStick], you have to be really, really careful.”} - \textbf{P10}
\end{quote}

\begin{figure}[t]
    \centering
    \includegraphics[width=0.95\linewidth]{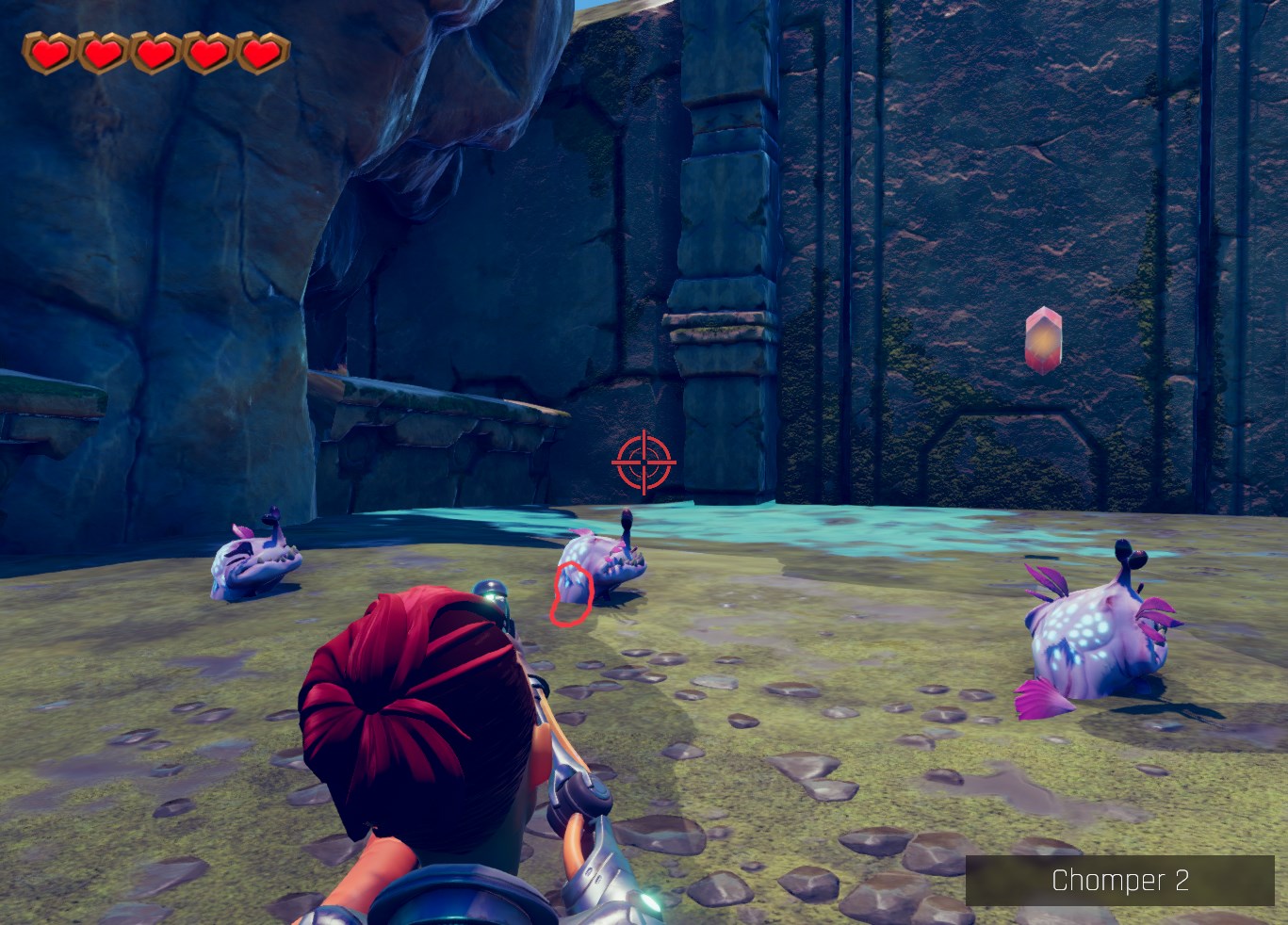}
    \caption{A still shot from \textit{The Explorer}, our prototype 3D adventure game. The main character (``Ellen'') is aiming a laser gun at an enemy (a "Chomper"). There are three Chompers in this view: one to the left, another to the right, and one in the middle under the crosshairs. An unlighted (pink) checkpoint crystal can be seen in the distance toward the right.}
    \label{fig:demo_game_ss}
\end{figure}

% \section{Study \replaced{2}{1}: \replaced{Understanding NavStick's Potential In Existing 3D Games }{Evaluating NavStick in Game Environments}}

\section{Study 2: Understanding NavStick's Potential In Existing 3D Games}

NavStick’s \added{ultimate} goal is to make \textit{existing} 3D video games --- ones that sighted players play --- more blind-accessible so that VIPs are not restricted to highly simplified games made just for them. Our core belief is that being able to ``look around,'' as NavStick affords, is integral to having an authentic experience while playing 3D games. In Study \replaced{2}{1}, we \added{complement Study 1 with an exploratory study where we }use \textit{The Explorer}, a representative 3D adventure game we created, to \replaced{understand NavStick's potential in making existing}{investigate the ways in which NavStick can make} 3D games more blind-accessible.

We investigate the following research questions in Study \replaced{2}{1}:
\begin{enumerate}[label=(RQ\replaced{2}{1}.\arabic*), leftmargin=5.1\parindent]
    \item Using \textit{The Explorer} as a case study, in what ways can NavStick make 3D games more blind-accessible?
    \item To what extent does NavStick work for complex situations found in mainstream 3D games --- namely, moving enemies, time pressure, and environmental occlusion?
\end{enumerate}

\subsection{Study 2 Game: \textit{The Explorer}}

\begin{table*}[ht]
\begin{tabular}{c|ccc|c}
\textbf{\#} & \textbf{M}                        & \textbf{TP}                       & \textbf{O}                        & \textbf{Objective}                                                                                                               \\ \hline
\textbf{1}          & -                                 & -                                 & -                                 & Find the CP behind 5 crates.                                                                                                              \\
\textbf{2}          & {\color[HTML]{009901} $\boldcheckmark$} & -                                 & -                                 & Defeat a R.\ Chomper to reveal the CP.                                                                                        \\
\textbf{3}          & -                                 & {\color[HTML]{009901} $\boldcheckmark$} & -                                 & Defeat 4 S. Chompers to reveal the CP. \textit{(60 sec.)}                                                                              \\
\textbf{4}          & -                                 & -                                 & {\color[HTML]{009901} $\boldcheckmark$} & Find a PP to open a gate, revealing the CP.                                                                           \\
\textbf{5}          & {\color[HTML]{009901} $\boldcheckmark$} & {\color[HTML]{009901} $\boldcheckmark$} & -                                 & Defeat a R.\ Chomper to reveal the CP. \textit{(45 sec.)}                                                                              \\
\textbf{6}          & {\color[HTML]{009901} $\boldcheckmark$} & -                                 & {\color[HTML]{009901} $\boldcheckmark$} & Traverse a grove containing 4 R.\ Chompers.                                                                                          \\
\textbf{7}          & -                                 & {\color[HTML]{009901} $\boldcheckmark$} & {\color[HTML]{009901} $\boldcheckmark$} & Traverse a grove containing 9 S.\ Chompers. \textit{(3 min.)}                                                                                  \\
\textbf{8}          & {\color[HTML]{009901} $\boldcheckmark$} & {\color[HTML]{009901} $\boldcheckmark$} & {\color[HTML]{009901} $\boldcheckmark$} & \begin{tabular}[c]{@{}c@{}}Find a PP behind 7 R.\ Chompers. CP beyond gate. \textit{(3 min.)}\end{tabular}
\end{tabular}
\vspace{2mm}
\caption{Overview of Study \replaced{2}{1} game segments. \textit{Conditions present:} M = enemy movement, TP = time pressure, O = occlusion. \textit{Other POIs:} PP = pressure pad, CP = checkpoint. \textit{Enemies:} R.\ Chomper = Roaming Chomper, S.\ Chomper = Stationary Chomper. Time limits are indicated in parentheses. \textbf{(Our supplementary material includes a visual overview of these segments with screenshots.)}}
\label{table:s1_overview}
\end{table*}

We created a 3D adventure video game called \textit{The Explorer}, derived from Unity Software Inc.’s 3D Game Kit~\cite{Unity3DGameKit2018}, to evaluate NavStick. We stress that \textit{The Explorer} is \textit{not} a game designed specifically for blind players, as most of today's audio games are. Rather, \textit{The Explorer} is a 3D video game with graphics, akin to what sighted players play, that we are simply ``attaching'' NavStick to in order to make it blind-accessible. \textit{The Explorer} is set within a fantasy world that the main character (``Ellen'') must traverse while using a laser gun to defeat enemies. Figures \ref{fig:navstick_overview}, \ref{fig:zoom_session}, and \ref{fig:demo_game_ss} show screenshots from \textit{The Explorer}. 

\textit{The Explorer} consists of eight segments (short levels). In each segment, players must traverse the area from a start checkpoint to an end checkpoint. We designed the eight segments to serve as a vehicle for investigating RQ2.2; namely, to investigate the extent that NavStick works for complex situations found in mainstream 3D games. To this end, we identified three environmental characteristics that can potentially present a challenge for NavStick --- enemy movement, time pressure, and environmental occlusions --- and designed the eight segments to feature every possible on/off combination ($2^3 = 8$ in total) of these characteristics.

Table \ref{table:s1_overview} summarizes the segments. Our supplementary material includes a visual overview of the segments with screenshots (see \texttt{VisualSegmentOverview.pdf}).

Segments 2, 5, 6, and 8 feature \textbf{enemy movement}. \textit{The Explorer} features two types of enemies: ``Roaming Chompers'' and ``Stationary Chompers.'' The Roaming Chomper continuously walks back-and-forth along a straight line, making a squeaking sound when it turns around. The Stationary Chomper stays at one position and does not move around. Segments with enemy movement feature one or more Roaming Chompers, and players must continuously track the Roaming Chompers' positions to aim their laser gun correctly.

Segments 3, 5, 7, and 8 feature \textbf{time pressure}. In these segments, we impose a pre-defined time limit, specified in Table \ref{table:s1_overview}. During these segments, soft clock-ticking sounds play in the background, and a voice announces the time remaining at regular intervals. If the player does not complete the segment in time, we restart it and allow them to try again. We allow a maximum of three tries before moving on.

Segments 4, 6, 7, and 8 feature \textbf{environmental occlusions}. In these segments, we placed trees in the environment that occlude NavStick's line-of-sight to POIs. Participants may need to move around the trees if they are not able to find a line-of-sight target using NavStick.

Players move Ellen with the left thumbstick. Tilting it forward and backward will allow the character to move forward and backward. Tilting it left and right will rotate the character left and right. NavStick is mapped to the right thumbstick as described previously.

The player's primary weapon is a laser gun, which they can aim and fire at enemies. The controls for aiming and firing are similar to those in mainstream action-adventure games such as \textit{Uncharted 4}~\cite{NaughtyDog2016} and \textit{Borderlands 3}~\cite{GearboxSoftware2019}. Holding the left trigger \textit{(LT)} enters ``aiming mode,'' in which the left thumbstick aims the gun rather than moves Ellen. Pressing the right trigger \textit{(RT)} while in aiming mode fires the gun. Only one hit is needed to defeat each enemy. Enemies do not attack the player.

NavStick considers enemies, checkpoints, pressure pads, and doors as POIs. All sounds in the game are spatialized, except for the time pressure sound effects. The game plays additional spatialized sound effects when the player defeats enemies, shoots a projectile and misses, or reaches the end checkpoint of a segment.

\subsubsection{Quality-of-Life Enhancements}
\label{sec:qol-enhancements}

\hfill \newline \added{We performed initial pilot tests with three research team members and four external individuals. One person was blind, and the others were sighted but wore blindfolds. These tests comprised an initial test of the study's procedure and}\deleted{Initial pilot tests} revealed that players needed supplementary tools to help them traverse \textit{The Explorer}'s segments. To this end, we added several quality-of-life enhancements to the game beyond NavStick. \deleted{Note that t}\added{T}hese additions are not intended to \textit{replace} NavStick but rather \textit{supplement} it. \added{(It is important to note that Study 1 \textit{did not} include these enhancements.)}

\textbf{Quest display}: Like other adventure games, \textit{The Explorer} features quest objectives such as reaching a target checkpoint that players must reach in order to proceed. VIPs can easily get lost without any indication of where they should be going. The quest display works to rectify this. It plays a short, spatialized tone every time the player turns $15\degree$. The tone originates from a fixed distance relative to the player and from the direction of the current target checkpoint. This tool serves two purposes: to give players a general sense of where their current target is located, and to help them understand their current orientation with respect to the target.

\textbf{Automatic vertical aim}: NavStick in its current form cannot scrub vertically, and its binding to the right thumbstick removes the player's ability to move the camera up and down. Because of this, enemies that are at a higher or lower elevation than the player cannot be hit by the laser gun since the laser travels straight ahead. We, thus, implemented an automatic vertical aiming functionality that automatically controls the vertical component of the player's aim. Section~\ref{sec:vertical-surveying} describes how future versions of NavStick can generalize to spherical surveying.

\textbf{Auto-turn}: Rotating Ellen to face a scrubbed target's direction can be quite challenging, so we implemented an “auto-turn” feature. Players can press the left bumper \textit{(LB)} while scrubbing to automatically turn Ellen toward the scrubbed target. The scrubbed target will be at the 12 o'clock position once the auto-turn is complete.

\textbf{Proximity notifier}: VIPs cannot see our game's graphics, and thus cannot tell whether or not they are near a POI. We implemented a “proximity notification” feature that automatically plays appropriate ambient sound effects while a player is physically near each type of POI.

\textbf{Increased collider sizes}: We doubled the collider dimensions around POIs to make them easier to target with NavStick.

\textbf{Footstep feedback}: Visually impaired pilot participants requested acoustic feedback on whether or not they are moving. As such, we added velocity-dependent footstep sounds that play as Ellen moves through the world.

\textbf{NavStick post-announcement tone}: When a player scrubs to a POI \added{(an enemy, checkpoint, pressure pad, or door)}, NavStick announces the name of the object and then plays a low-pitched (440\,Hz) sustained sine tone as long as the player keeps pointing at the object. This allows players to continuously track objects such as Roaming Chompers.

\textbf{Non-POI tone}: By default, NavStick does not say anything if it is \textit{not} being pointed at a POI. However, this may cause the world to sound ``empty.'' Thus, if a player scrubs in a direction where the first object in that direction is \textit{not} a known POI (e.g., a wall or tree), NavStick will immediately play a higher-pitched (1,320\,Hz) sustained sine tone originating from the point of the obstruction in space (with \textit{no} verbal announcement).

\subsubsection{\added{Limitations}}

\hfill \newline \added{NavStick's goal is to give VIPs the ability to look around within a game world, giving them \textit{more} of the type of experience that sighted players have when playing 3D games. In our initial pilot tests of NavStick (Section \ref{sec:qol-enhancements}), we found that NavStick by itself cannot make a 3D game entirely blind-accessible --- there are times in which players need other mechanics as well, such as an indicator showing players what general direction their goal is (in our case, the quest display) and a means of aiming vertically (in our case, automatic vertical aim). These quality-of-life enhancements may not be available in every game, preventing our results from being applicable to \textit{any} 3D game without some degree of modification. Nevertheless, by integrating these enhancements, we were able to learn about what NavStick enables and what it does not enable by itself, and in Study 2, we explore what NavStick can enable VIPs to do when it is complemented by these enhancements.}

% We also believe that NavStick must also work alongside other mechanics (here, the quality-of-life enhancements) in order to be successful. However, the quality-of-life enhancements we implemented in \textit{The Explorer} may not be available in every game --- this may prevent our results for Study 2 from being applicable to \textit{any} 3D game without some modification. Nevertheless, our goal with this study is to explore NavStick's \textit{potential} in making existing 3D games blind-accessible, seeking to understand how VIPs receive NavStick in the context of a representative 3D adventure game.}

\begin{figure}[]
  \centering
  \includegraphics[width=0.95\linewidth]{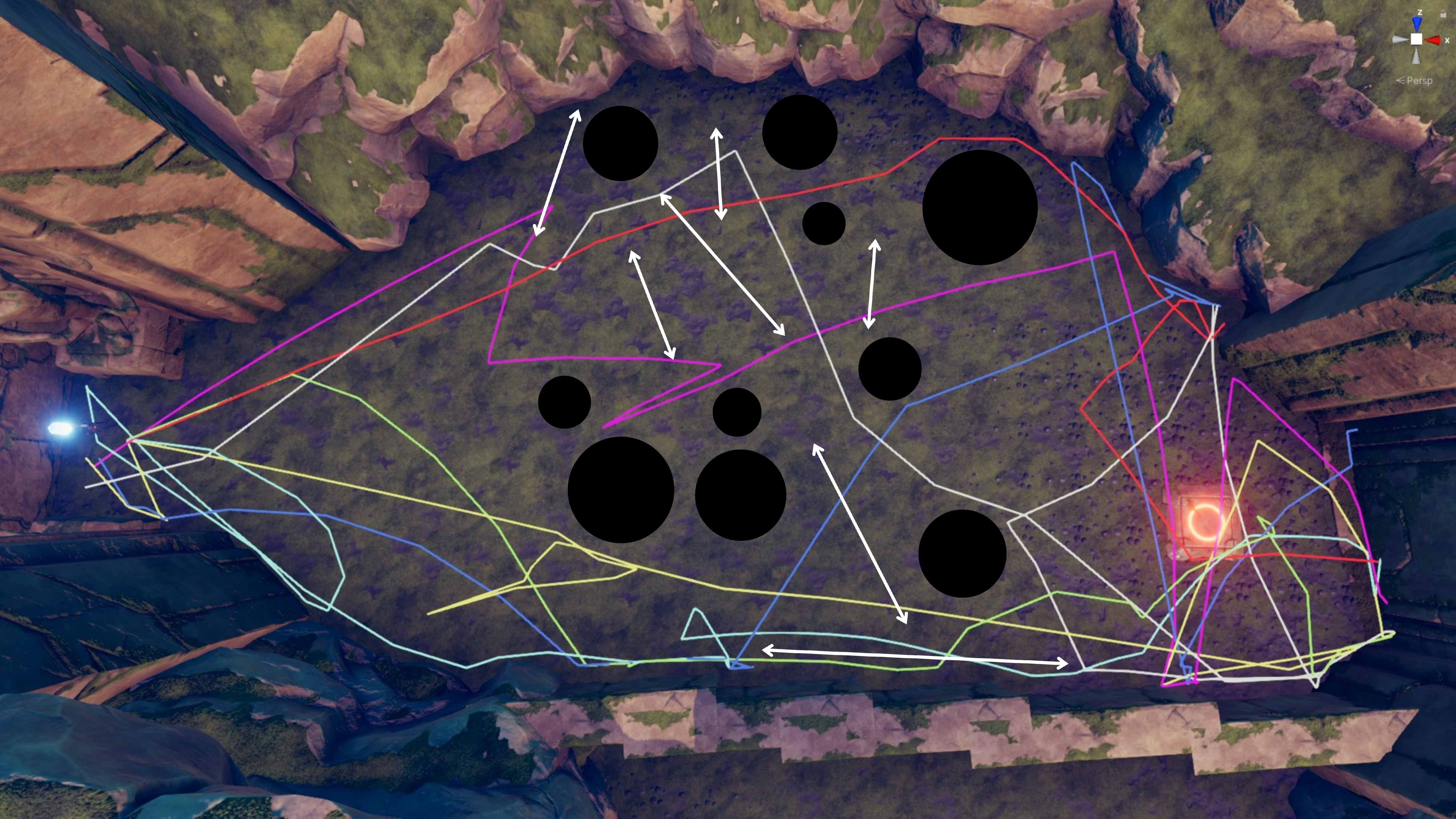}
  \caption{Player movement trails within Segment 8. Players start at the glowing checkpoint (far left) and must reach the glowing pressure pad (lower-right) to open a large gate (partially visible on far right). Each color represents a different player's path. We can see that NavStick afforded these players a high degree of agency in how they played this segment. Trees are shown as black circles for clarity; these act as occluders and cannot be walked or seen through. White line segments represent the seven Roaming Chompers' patrol paths. No two players defeated the same set of Chompers.}
  \label{fig:segment_8_trails}
\end{figure}

\subsection{Study 2 Participants}

Study \replaced{2}{1} involved seven participants, six of whom described themselves as having no usable vision and one (P1) who described themselves as "otherwise visually impaired." \added{Five had participated in Study 1.} All participants were male and have had their vision impairments from birth. Five participants (P1--5) were 18--25 years old; two participants (P6 and P7) were 26-35 years old. Two participants (P4 and P7) reported minor hearing loss in their right ear. All but one participant (P4) reported themselves to be experienced with video games (4+ on a 5-point Likert scale). \replaced{We recruited these participants in the same way as Study 1.}{We recruited participants by posting to online forums popular among the visually impaired community \textit{(exact forums anonymized for submission)}.}

\subsection{Study 2 Procedure}

We began by administering a pre-study questionnaire requesting demographic information. We then guided players through three trial levels as we explained the game's mechanics. Trial Level 1 introduced pressure pads, doors, and checkpoints as well as NavStick itself. Trial Level 2 introduced a Stationary Chomper and the laser gun. Trial Level 3 introduced a Roaming Chomper. 

We began each of the eight segments by explaining the segment's objective (Table~\ref{table:s1_overview}). After the player completed each segment, we paused the game and administered a post-segment questionnaire. The first part gauged participants’ subjective measures of effort, frustration, fun, and impression in the segment using 20-point Likert scales similar to a NASA TLX form~\cite{Hart1988}. In the second part, we revealed the environmental characteristic(s) that the segment possessed (the checkmarks in Table~\ref{table:s1_overview}) and asked participants to reflect on how NavStick performed with respect to those characteristics.

At the end of the session, we administered a post-study questionnaire and a semi-structured interview to ask participants to reflect on what they liked and disliked about NavStick as well as their impression of the game itself. 

\subsection{Study 2 Results}

Below, we report the main takeaways from Study \replaced{2}{1} in answering our research questions: learning the ways in which NavStick can make 3D games more blind-accessible (RQ\replaced{2}{1}.1), and learning how well NavStick performs in the face of enemy movement, time pressure, and occlusion (RQ\replaced{2}{1}.2).

We analyzed our qualitative results (participant quotes\deleted{ and Likert scale results}) \replaced{identically to Study 1.}{via qualitative coding. All authors collectively clustered data into common themes, debating among themselves until there was mutual agreement on the themes. Each theme presented below represents opinions shared by a majority of our participants.} \added{Any quantitative data (Likert scale results) were analyzed as described in the following subsections.}

\subsubsection{RQ 2.1 Results: Making 3D Games More Blind-Accessible}

\hfill\newline All participants were able to complete all eight segments of \textit{The Explorer} using NavStick to look around. In the timed segments, participants ran out of time in only three instances: P3 and P4 in Segment 8 and P5 in Segment 5. P3 and P4 did not expect the Segment 8 pressure pad to be close to the gate (they were far away from each other in Segment 4); P5 lost track of the Segment 5 checkpoint and forgot about using the quest display to reorient themselves. However, all three participants were able to complete the segments on their second try.

In addition to making \textit{The Explorer} merely possible to play, we found evidence that NavStick afforded players agency; that is, letting them pursue their own strategies while playing. In-game analytics data from Segments 7 and 8 show that no two participants defeated the same batch of Chompers in either segment. Furthermore, no participant defeated all of the Chompers in these segments; they defeated some subset and decided to move on. Figure \ref{fig:segment_8_trails} shows the paths that participants took within Segment 8 --- note how participants took many unique paths to get to the target checkpoint. Also note how the trails become random around the pressure pad --- once players hit the pressure pad, many of them freely roamed around it while the gate opened. Some participants noted this freedom:

\begin{quote}
    \textit{“I loved that there’s strategy. It depends how you want to play it. I enjoy stuff like that. Do I want to let [the Roaming Chomper] come to me? Do I want to proactively hunt it? I like that. That’s what I’m saying: This [type of gameplay] should be in a game.”} - \textbf{P6}
\end{quote}

The last sentence of this quote sheds light on how inequitable the current world of video games is for VIPs. This type of gameplay is \textit{already} present within video games made for sighted players. However, due to the gap between games for sighted players and blind-accessible games, our participants were not used to playing a game that promotes such a high level of freedom and requires such a high level of strategy.

Five of the seven participants wished that NavStick was available in existing games, explicitly mentioning \textit{The Last of Us Part II}~\cite{NaughtyDog2020} and \textit{Halo 2}~\cite{Bungie2004} as candidates for NavStick. The former is especially interesting given that it has \textit{many} accessibility features~\cite{NaughtyDogSonyEnt2020} and has received many accolades for them~\cite{MolloyCarter2020, Watts2021}. Participants, however, agreed that NavStick offers much more freedom than the guidance-based tools that \textit{The Last of Us Part II} provides. They added that NavStick promoted ``seeing'' in the form of providing an \textit{understanding} of the environment:

\begin{quote}
    \textit{“I mean, if you stop to think about it, in most games, if you’ve ever played first-person, the whole point is that you can move the camera. And to me, [NavStick] was the equivalent of a camera. You’re able to see objects around you. You’re able to know where they are.”} - \textbf{P6}
\end{quote}

By providing the equivalent of a camera for VIPs to look around, NavStick was able to provide a sense of freedom and environmental understanding that VIPs do not currently experience in existing games. We found this quote interesting because NavStick repurposes the game controller's right thumbstick, which is used by sighted players to control the game's camera.

Lastly, four particpants reported having fun in a way that they are not used to in existing games:

\begin{quote}
    \textit{“It was very enjoyable. It’s very fun to look around your environment, figure out where you’re going, figure out what you have to do, and to blow stuff up.”} - \textbf{P2}
\end{quote}

This quote is insightful because it aligns with established definitions of player agency from the game development community (using the synonym \textit{intention})~\cite{Church1999a, Church1999b, Church2006} and from the computer science community~\cite{Yuan2011}. Specifically, it highlights how NavStick enables the three activities critical to agency: understanding the environment, developing a plan of action, and attempting to execute that plan.

%%%

\begin{figure}[t]
    \centering
    \includegraphics[width=0.95\linewidth]{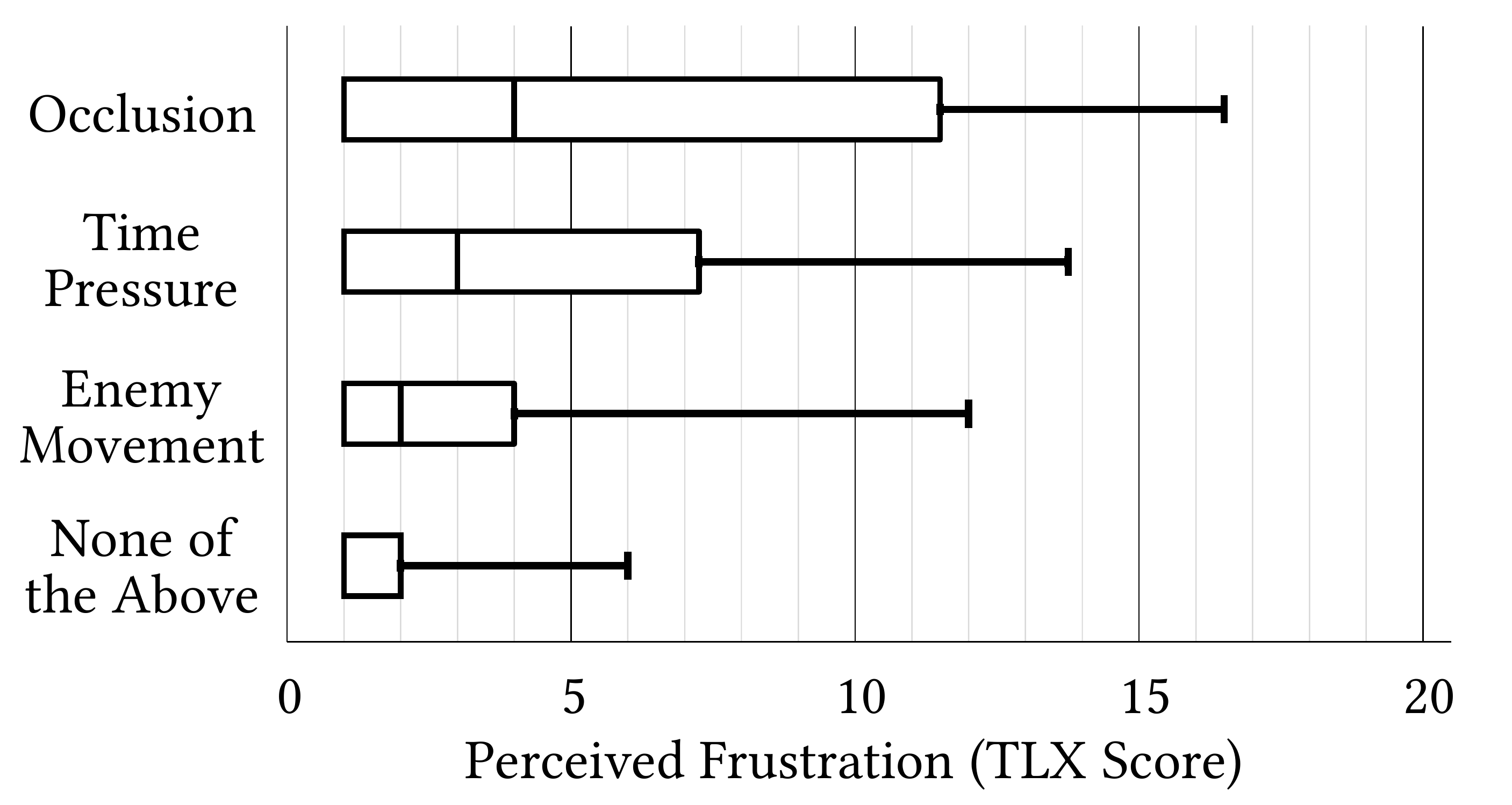}
    \caption{Box plot indicating distribution of TLX frustration scores across segments that featured occlusion, time pressure, enemy movement, and none of the three.}
    \label{fig:s1_frustration_chart}
\end{figure}

\subsubsection{RQ 2.2 Results: Enemy Movement, Time Pressure, and Occlusion}

\hfill\newline Here, we present our findings on how well NavStick performs in the face of enemy movement, time pressure, and environmental occlusion. Figure \ref{fig:s1_frustration_chart} shows a box plot of TLX frustration scores across segments that featured each of these characteristics. Our analysis revealed that participants found segments with occlusion ($M = 5.8$, $SD = 5.0$) to be the most frustrating overall, followed by segments with time pressure ($M = 4.3$, $SD = 3.9$), enemy movement ($M = 3.4$, $SD = 3.3$), and none of the three characteristics ($M = 2.3$, $SD = 2.1$), in that order. This result was affirmed by participants' comments during our semi-structured interviews, in which most remarked that occlusion was the most difficult factor to deal with when using NavStick.   

NavStick communicates occlusions by rendering a high-pitched sine tone at the raycast's point of impact whenever that point is not an actual target. Participants remarked that this was helpful, but five out of seven participants felt that they needed more information than this, specifically a description of the obstruction and its dimensions:

\begin{quote}
    \textit{"I will say that it’s not clear what’s blocking me. Is it something I could fly or jump over? [...] I couldn’t tell what [exactly] it was either, so going around it is a bit difficult when you don’t know what the thing is you’re trying to go around.”} - \textbf{P2}
\end{quote}

% Any attempt at integrating such data alongside NavStick, however, should take care to not overload the user with information.

Participants had differing opinions on how well NavStick handled time pressure and enemy movement. With respect to time pressure, four participants felt that NavStick allowed them to survey the area quickly:

\begin{quote}
    \textit{“This is what Navstick does especially well. It allows me to really quickly scan around the area. The time restriction made me really need the ability to scan around quickly.”} - \textbf{P7}
\end{quote}

The other three participants, however, felt that the speed required in a timed situation was incompatible with NavStick’s behavior of truncating speech when moving off a slice:

\begin{quote}
    \textit{“The problem with NavStick is that when I move just a little bit, the voice indicating what is in front of me stops, [and] I have to re-move the stick again. So it has a pretty small target area [and] it made it pretty difficult to use because of that.”} - \textbf{P5}
\end{quote}

As mentioned previously, this aspect of NavStick follows earPod's design~\cite{Zhao2007}. Its authors found that truncated audio playback gave users  a much stronger feeling of responsiveness and was less confusing to use than simultaneous playback (where announcements are always played in their entirety). Our findings here, however, suggest that it may be necessary to allow for some ``residual'' information about the previous slice to remain after the player scrubs away from it, which is an expansion of the findings from earPod.

Participants also had differing opinions with respect to enemy movement. Roaming Chompers emit spatialized footstep sounds that are audible when the player is sufficiently close, but participants were split on whether NavStick was needed if enemies' footsteps were already spatially audible. Some believed that NavStick offered a useful supplement to the sound effects' spatialized audio:

\begin{quote}
    \textit{“I couldn’t immediately hear where [the Roaming Chomper] was. I knew it was near me. [...] But I didn’t know where it was until I used [NavStick] to locate it.”} - \textbf{P2}
\end{quote}

Others, however, felt that NavStick was unnecessary in these situations, since these players could tell which way the Roaming Chompers were moving from their footstep sounds alone:

\begin{quote}
    \textit{“The Chomper already makes noise, so [using NavStick is] just extra effort and brain processing power. It’s like I’m trying to align NavStick with something I can already hear and there’s not really a benefit to it.”} - \textbf{P7}
\end{quote}

The difference between these groups may be that the latter group is already accustomed to discovering and monitoring POIs (such as enemies) purely using sound. P4, P5, and P7, who comprise the latter group, went on to describe their experiences with existing audio-based games such as \textit{Swamp}~\cite{Kaldobsky2011} and \textit{BK3}~\cite{NyanchanGames}. These games announce the presence of objects by playing relevant ambient sound effects when the player gets near them.

We should note, however, that NavStick gives players the choice of whether to use it or not. \added{That is, players could use it as much or as little as they wanted.} \replaced{Those}{Players} who want or need to use NavStick to locate objects can do so, and \replaced{those}{players} who prefer to locate objects from their ambient sound effects alone can simply not use NavStick for that purpose. NavStick's on-demand nature means that it will stay out of the way if the player does not wish to use it and is ready for whenever they \textit{do} require it.

\section{Implications for Blind-Accessible Games}

The findings from our two studies yielded several implications for future blind-accessible games. Here, we reflect on the findings and propose ideas for next steps.

\comment{Re-ordered the sections due to the change in ordering of the studies.}

\subsection{NavStick and menu-based surveying should co-exist}

In forced rankings within Study \replaced{1}{2} (see Figure \ref{fig:pref_chart}), participants strongly preferred NavStick for navigation tasks that rely on having a mental map of the surrounding environment (both ``directional'' tasks), as well as for navigation within video games (the \textit{Terraformers} task). However, for other navigation tasks, their preference was split between NavStick and NavMenu. Players viewed the two tools as working hand-in-hand with each other. The ability to be efficiently \textit{guided} toward POIs in the environment (as represented by NavMenu) and the ability to \textit{look around} (as represented by NavStick) ultimately serve different purposes and should work in tandem with one another.

As a result, we believe that future blind-accessible games and virtual navigation tools should offer \textit{both} the ability to look around (as offered by NavStick) and guided navigation (as offered by existing audio navigation systems) so that users can choose which they want to use at any given moment.

\subsection{NavStick can generalize to spherical and real-world surveying}
\label{sec:vertical-surveying}

In its current state, NavStick only operates in the user's horizontal direction. To facilitate exploration in three dimensions, future versions of NavStick should allow users to scrub vertically, possibly by using a game controller's gyroscope to control pitch. Games such as \textit{Splatoon}~\cite{Nintendo2015} and \textit{The Legend of Zelda: Breath of the Wild}~\cite{Nintendo2017} already use gyroscopic controls for aiming vertically. This would allow a player to, for example, \added{scrub to high or low portions of grocery store shelves (like the ones in Study 1) or even} scrub for objects on top of hills or other high points (within environments similar to \textit{The Explorer})\deleted{ or even scrub to high or low portions of grocery store shelves (like the ones in Study 2)}. It may also facilitate looking around in all directions within immersive mixed reality environments.

Although we developed NavStick for virtual environments, we believe that this new approach could also apply to physically navigating in the real world. It is important to note that the real world brings new challenges, such as accuracy, precise tracking, and guaranteeing a user’s safety. Still, there is promise in adding NavStick's scrubbing-based interaction technique to existing physical devices beyond game controllers to facilitate real-world navigation. Future researchers may consider adding a thumbstick to a white cane, for example, or letting VIPs scrub their finger on the perimeter of a smartwatch face so they can survey and explore the real world more independently.

\subsection{Blind-accessible 3D games are possible, but more perceptual challenges await}

\replaced{Both studies showed that NavStick has the potential to make 3D video games more blind-accessible and offer VIPs a more fun and fulfilling experience in gaming. In Study 1, we found that NavStick allowed players a better sense of the space they were in compared to status quo tools for VIPs. In Study 2, }{Study 1 showed that making 3D video games blind-accessible is indeed possible:}\textit{all} participants completed all segments of \textit{The Explorer}, a representative 3D adventure game created for sighted players. To some of our participants, completing a game made for sighted players came as a surprise and felt different from blind-accessible games that they have played before. Indeed, one said that this was the first game where they could “just walk around and blow stuff up.” These results are encouraging for the prospect of making mainstream 3D games that sighted players play --- games such as \textit{The Legend of Zelda} --- blind-accessible in the future. 

We believe from our findings, however, that two significant perceptual challenges must still be overcome to give VIPs a fully equivalent experience playing video games. The first challenge is communicating the layouts of environments and the obstructions within them, and the second challenge is to communicate more of the environment's aesthetics to players. Both are related to participants' deep desire to ``see'' \textit{even more}.

Regarding the first challenge, many participants desired some form of ``x-ray vision'' --- the ability to see the world \textit{beyond} their immediate line-of-sight. At its core, NavStick revolves around the principle of replicating line-of-sight, but sighted players have access to a world map or minimap that indicates the shape of the area they are in and routes to adjacent areas. As a result, an important future step for blind-accessible 3D game worlds will be to give players an ``overview'' sense of a space beyond what they can survey with line-of-sight, especially if future systems allowed players to survey portions of the game world on demand with these techniques.

Regarding the second challenge, participants wanted to know what the game environments looked like beyond merely knowing what objects they contained. 3D game worlds are often richly detailed and fantastical, and \textit{The Explorer} was no exception. The game design community characterizes \textit{sensation} as a major type of fun that games provide~\cite{Hunicke2004,LeBlanc2008}. We believe that allowing VIPs to appreciate the aesthetics of a 3D world is crucial to making 3D games \textit{equivalently accessible} to VIPs. Sensory substitution techniques~\cite{Hamilton-Fletcher2016} and automatic image captioning techniques~\cite{Dognin2020} are two promising ways of communicating aesthetics since they can translate images to sound or textual descriptions. Future systems might combine one of these approaches with NavStick-style scrubbing to give players aesthetic information in a self-directed way, allowing them to answer the question, ``What does the environment in \textit{this direction} look like?"

%%%

\section{Conclusion and Limitations}

In this work, we introduce NavStick, an audio-based tool for looking around within virtual environments. Its aim is to make 3D video games --- ones designed for sighted players --- more blind-accessible by fostering agency in navigation. NavStick repurposes a game controller’s thumbstick to allow VIPs to survey what is around them via line-of-sight.

Through two\deleted{ extensive user} studies, \replaced{we investigated NavStick with respect to status quo methods for surveying a game world (Study 1) and explored NavStick's potential in making existing 3D video games more blind-accessible (Study 2).}{we investigate NavStick with respect to both game environments (Study 1) and navigation tasks (Study 2). We also compare NavStick with existing menu-based surveying techniques (Study 2).} \replaced{We found that VIPs were able to form more accurate mental maps of their environment with NavStick than with menu-based surveying; we found that players wanted the two types of surveying to co-exist; w}{W}e found that NavStick offers players a sense of agency not found in current blind-accessible games; \added{and} we revealed new challenges that must be overcome to make 3D games fully accessible to VIPs\replaced{.}{; we found that VIPs were able to form more accurate mental maps of their environment with NavStick than with menu-based surveying; and we found that players wanted the two types of surveying to co-exist.}

As with many studies that involve VIPs, we had a relatively low number of participants. Although we found several significant results through our surveys and interviews, there may be more insights that our user studies could not reveal regarding how NavStick could better integrate into a video game or how it compares to menu-based surveying tools. In addition, while we are proud of the fact that our research team included VIPs as co-designers throughout the process, we regret not being able to recruit a more diverse group of visually impaired study participants.

We hope that, just as sighted people are able to look around and explore complex and rich game environments on their own terms, so too can visually impaired people. Whether in a single-player adventure game or a multiplayer role-playing game, it is our hope that visually impaired players can one day play the same games as sighted players in an equivalent fashion.

\begin{acks}
We would like to thank Brian (Shao-en) Ma and Emily Li for their assistance in preparing for and executing Study 2. We would also like to extend our sincere gratitude toward our study participants for their cooperation and for providing us with helpful feedback. This work was supported in part by a Junior Faculty Grant from Columbia University's Office of the Provost.
\end{acks}

\bibliographystyle{ACM-Reference-Format}
\bibliography{sample-base}

\end{document}